%% file: omega.tex
\documentclass [a4paper, 12pt]{revtex4-1}

\usepackage{amsmath,amssymb}
\usepackage{epsfig}
\usepackage{graphicx}
\usepackage{multirow}
\usepackage{color}

\begin{document}

\title{Precise Omega baryons from lattice QCD}

\author{Renwick~J.~Hudspith}
\affiliation{GSI Helmholtzzentrum f\"ur Schwerionenforschung, 64291 Darmstadt, Germany}

\author{Matthias F.M. Lutz}
\affiliation{GSI Helmholtzzentrum f\"ur Schwerionenforschung, 64291 Darmstadt, Germany}

\author{Daniel Mohler}
\affiliation{Institut f\"ur Kernphysik, Technische Universit\"at Darmstadt, Schlossgartenstra{\ss}e 2, 64289 Darmstadt, Germany}
\affiliation{GSI Helmholtzzentrum f\"ur Schwerionenforschung, 64291 Darmstadt, Germany\\}

\begin{abstract}
In this paper we determine the masses of $I(J^P)=0\left(3/2^+\right)$ and $0\left(3/2^-\right)$ $\Omega$-baryon ground states using lattice QCD. We utilise Wilson-clover ensembles with $2+1$ dynamical quark flavours generated by the CLS consortium along a trajectory with a constant trace of the quark-mass matrix. We show that N$^3$LO $\text{SU}(3)_f$ chiral perturbation theory expressions describe the ground-state masses with positive-parity well, and we use them to set the lattice scale. Methodologically, our combination of gauge-fixed wall sources and the generalized Pencil of Functions allows for high-precision determinations of the lattice spacing at a relative error of around $0.3\%$ with controlled excited-state contamination. The fit we perform allows for the continuum value of $t_0$ to vary, thereby determining this quantity with a comparable level of precision to that of the lattice scale. Using the resulting scales our measurement of the negative-parity $\Omega^{3/2^{-}}$ state is found to be consistent with the recently-discovered $\Omega(2012)^-$, which can therefore be assigned the quantum numbers $I(J^P)=0\left(3/2^-\right)$.
\end{abstract}

\date{\today}
\maketitle

\input{introduction}
\input{lattice_methodology}
\input{fitform}
\input{results}
\input{results_fits}
\input{conclusions}

\section*{Acknowledgements}

The authors would like to acknowledge useful discussions with Konstantin Ottnad and Simon Kuberski. In addition the authors thank Yonggoo Heo for cross-checking some of the expressions in Appendix A. RJH would particularly like to thank Hartmut Wittig for inspiring the completion of this project. We thank the CLS consortium for providing gauge conﬁgurations. D.M. acknowledges funding by the Heisenberg Programme of the Deutsche Forschungsgemein-
schaft (DFG, German Research Foundation) – project number 454605793. 
Calculations for
this project were partly performed on the HPC cluster “Mogon II” at JGU Mainz. This research was supported in part by the cluster computing resource provided by the IT Division at the GSI Helmholtzzentrum f\"ur Schwerionenforschung, Darmstadt, Germany (HPC cluster Virgo).  For the light- and strange-quark propagator inversions we used the version 1.6 of OpenQCD \cite{Luscher:2012av}, QDP++ \cite{Edwards:2004sx}, and a multi-grid solver implemented atop GRID \cite{Boyle:2016lbp}. For the Coulomb gauge fixing and the independent measurements of $t_0$ we used the library GLU \cite{Hudspith:2014oja}. The GNU Scientific Library (GSL) \cite{Galassi2018} was used for the analysis in this work.

\appendix

\input{appendices}

\bibliographystyle{apsrev4-1}

\bibliography{omega}

\end{document}

%% file: introduction.tex
\section{Introduction}

Lattice quantum chromodynamics (lattice QCD) is now firmly established as the method of choice for calculations of the strong interaction in the non-perturbative regime, and the overall statistical resolution of lattice-determined quantities of interest is continuously improving. An example of such a quantity is the leading order hadronic contribution to the magnetic moment of the muon $((g-2)_\mu)$ \cite{Aoyama:2020ynm}, where uncertainties of QCD determinations are falling below $1\%$ \cite{Borsanyi:2020mff}.

In this ``precision era" of Lattice QCD, sub-$0.5\%$ determinations of the scale in lattice simulations are of paramount importance to high-precision measurements, as they are vital to make sure the scale-setting isn't the dominant source of uncertainty in the calculation \cite{DellaMorte:2017dyu}. In a minimal sense, for an $n_f=2+1$-flavor simulation, 3 physical quantities need to be used to determine the ``physical point". These are thereby sacrificed as quantities that can be predicted by the theory. Typically, the iso-symmetric pion and kaon masses are used, plus a third quantity that sets the overall physical scale. In short, this procedure is referred to as ``setting the scale".

Typically one chooses a particularly simple-to-measure physical quantity, which, together with the pion and kaon masses, then defines a ``renormalisation trajectory'' that sets the approach to the continuum. Of course any such choice has its own discretisation effects, so lattice spacing determinations from different quantities must differ. Historically, popular options have been the pion/kaon decay constants \cite{FermilabLattice:2011njy,Bruno:2016plf} or masses of light or light-strange baryons e.g. the Nucleon \cite{Bali:2012qs}, the Cascade \cite{RQCD:2022xux}, or the fully-strange Omega baryon mass $M_\Omega$ \cite{RBC:2014ntl,Miller:2020evg}. Even the splitting between the $\Upsilon^\prime$ and $\Upsilon$ in lattice NRQCD has been used in the past \cite{HPQCD:2003rsu}. Nowadays, an intermediate scale such as $r_0$ \cite{Necco:2001xg} or $r_1$ \cite{Bernard:2000gd}, $t_0$ \cite{Luscher:2010iy}, or $w_0$ \cite{BMW:2012hcm} is often used, although its value in the continuum must still be inferred from another experimentally-measurable quantity. All of the determinations of the lattice scale have their own challenges -- both in measurement, and regarding their associated systematic uncertainties.

It is also possible to use suitable sets of physical quantities instead of a single observable. For instance, in \cite{Lutz:2023xpi} the baryon octet and decuplet baryons masses were used simultaneously to set the scale. Whatever states, or combinations thereof, are used in the determination of the scale no longer are predictions of the theory. This makes quantities like the kaon decay constant $f_K$, for which there is a phenomenological interest for tests of CKM-unitarity, undesirable as an observable for scale-setting. In this paper we will demonstrate that the $I(J^P)=0(3/2)^+$ $\Omega$-baryon ground-state mass $M_\Omega$ is relatively straightforward to obtain with good precision. Subsequently the measured $M_\Omega$ can be used in combination with dimensionless quantities based on the light meson masses and $t_0$ to set the scale. 

The $\Omega$-baryon provides a very enticing state for setting the scale as strange quark propagators are relatively cheap to compute in comparison to those of the light flavors. The noise to signal ratio is not large, no complicated improved currents and renormalisation is needed, and finite-volume effects are expected to be small. The $\Omega$ in pure QCD is stable and a determination of the mass in lattice QCD is therefore unambiguous. As the $\Omega$ has isospin 0, strong isospin-breaking effects appear in the sea rather than in the valence sector and are expected to be negligible; on top of this pure-QED effects due to the charged nature of the strange quarks are present but expected to be small \cite{CSSM:2019jmq}. The $\Omega$ is very precisely known experimentally too, making it an ideal candidate for scale-setting. With this in mind, we will present a highly-accurate lattice-spacing and $\sqrt{t_0}$ determination using $M_\Omega$ determined on the Coordinated Lattice Simulations (CLS) ensembles \cite{Bruno:2014jqa, Bali:2016umi}.

In the next section we will introduce our methodology, followed by a discussion of flavor-$\text{SU}(3)$ ($\text{SU}(3)_f$) chiral fits in Section~\ref{chiral}. We then proceed to present our $\Omega$-baryon masses in Section~\ref{sec:latres1} and the resulting determination of the lattice scale in Section~\ref{sec:latfits}. We also compare our determination to  previous ones using some of the same ensembles along different renormalisation trajectories \cite{Bruno:2016plf,RQCD:2022xux,Lutz:2023xpi}. In addition, we use our determined lattice scales to compute the physical mass of the negative parity $M_\Omega^{(3/2)^-}$-baryon and present the results of this in Section~\ref{sec:32minus}. The calculated mass is consistent with the mass of the recently-discovered $\Omega(2012)^-$ \cite{PhysRevLett.121.052003}.

%% file: lattice_methodology.tex
\section{Methodology}

We use the simple local operator for the $\Omega$ (lower color, upper spin indices) with open spin index $\kappa$,
\begin{equation}
  \Omega^{\kappa}_i(x) = \epsilon_{abc} (s_a^T C\gamma_i s_b) s^\kappa_c(x),
\end{equation}
focusing only on the polarisation-diagonal components \cite{RBC:2014ntl} to make the contraction particularly cheap. The parity of choice of choice is projected-out in the usual manner,
\begin{equation}
C_\Omega^{\pm}(t-t^\prime ) = \frac{1}{2}(1\pm \gamma_t)^{\kappa{\kappa^\prime}}\frac{1}{3}\sum_{i=x,y,z} \sum_x^{L^3} \Omega^\kappa_i(x,t)(\Omega^{{\kappa^\prime}}_i(0,t^\prime ))^{\dagger}\,.
\end{equation}
 We perform an average over the forward and backward propagating states \footnote{With particular care being taken when going through the temporal boundaries as anti-periodic boundary conditions are applied on the fermions in the solve.}, with the appropriate parity projections to obtain a correlator with the correct $J^P=3/2^\pm$ quantum numbers. These forward and backward states are measured to be statistically quite independent and come only at the cost of more contractions (at least for the periodic gauge field ensembles).

To ameliorate the cost of the creation of the deflated subspace used in our solver \cite{Luscher:2007se} we will do as many solves as beneficial on each individual configuration. Here we invoke the Truncated Solver method \cite{Bali:2009hu}: For periodic ensembles in time we use translational-invariance in time and perform low-precision (sloppy) inversions on every timeslice $(t^\prime)$, and one exact solve randomly chosen in the bulk. For the open boundaries we start sufficiently far from the boundary and average every forward-propagating correlator on every timeslice from there until a cutoff below $L_t/2$ and likewise the backward propagating opposite-parity projected correlator from $L_t-t^\prime$ decreasing the source position. This also makes use of translational-invariance in time but restricted to the bulk away from the boundaries. We  measure the negative-parity, $J^P=3/2^-$ state in a similar fashion, as it is only a trivial parity-projection of the basic contracted correlator.

The one-end-trick (e.g. \cite{Foster:1998vw}) is not applicable to baryon correlators and simple generalizations suffer from a signal to noise problem as explained in \cite{ETM:2008zte}. Popular alternative choices are some form of point sources (with or without smearing) or Coulomb gauge-fixed wall sources \cite{Billoire:1985yn,Gupta:1990mr}. We will apply the latter option, which obviously requires us to fix the gauge, but yields a far superior volume-average. These have the limitation that the quarks carry a definite momentum but as everything we will consider will be at rest, thereby being no limitation for this work. Fixing to Coulomb gauge to high precision can be a numerically-cheap procedure if done using the Fourier-accelerated non-linear conjugate gradient algorithm of \cite{Hudspith:2014oja}, where an average gauge-fixing accuracy of $\Theta=10^{-14}$ can be achieved with negligible resources even for the largest volumes considered here ($96^3\times 192$). The main mass-trajectory CLS uses keeps the trace of the quark-mass matrix constant. As the pion mass decreases the strange quark is made heavier, simultaneously the volume is enlarged to ensure $m_\pi L \gtrapprox 4$. Hence, at the physical pion-mass point, the ensembles are expected to provide the most precise data and to most strongly constrain our extrapolations. 

\begin{figure}[h!]
\includegraphics[scale=0.45]{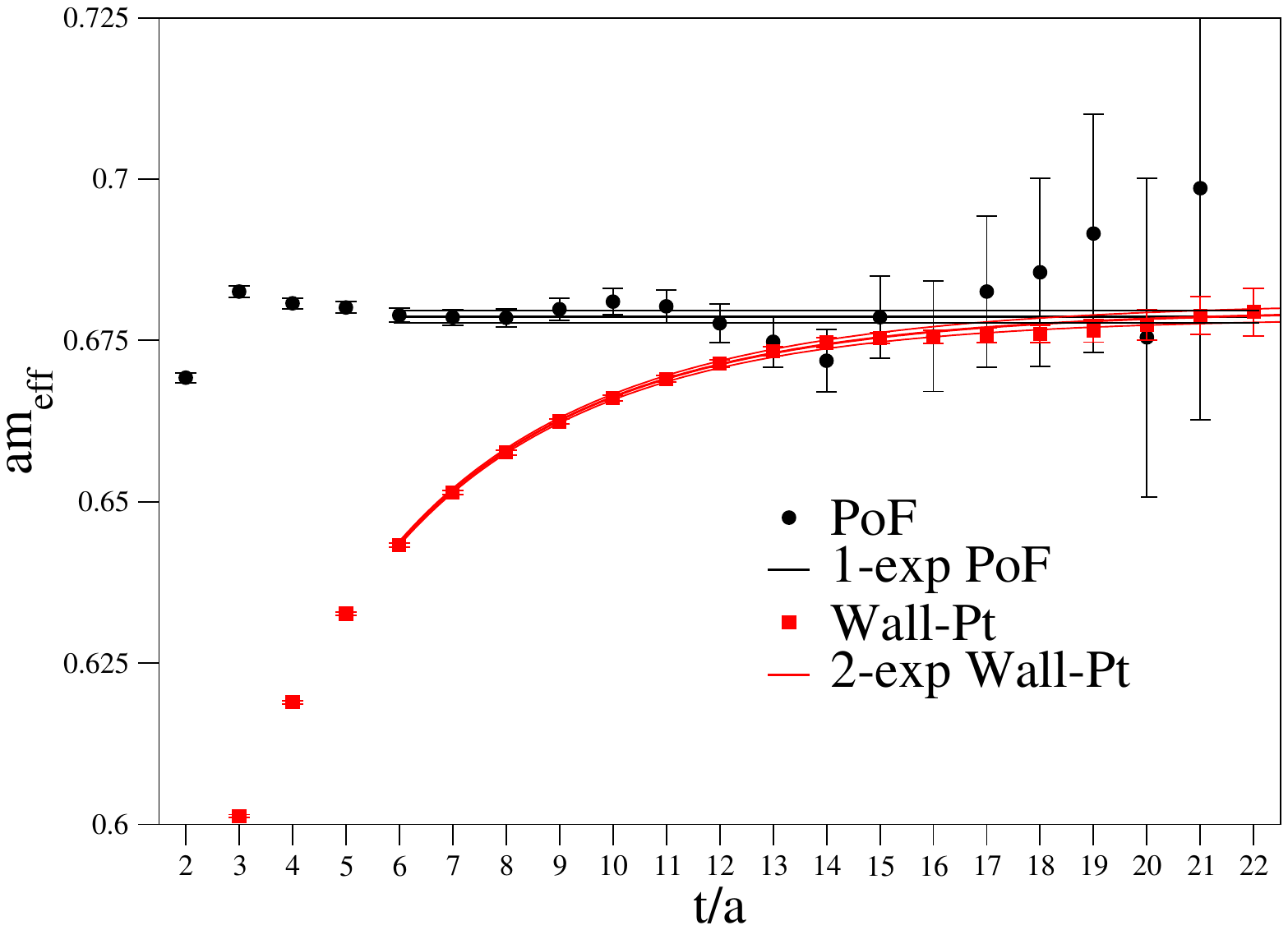}
\caption{A comparison of the lowest principal PoF- and the Wall-point correlators' effective masses of the positive-parity $\Omega$, for the ensemble C101.}\label{fig:C101_pof}
\end{figure}

As gauge-fixed wall sources approach the plateau region from below for the $\Omega$ and many other states beside pions (they are maximally asymmetric at source and sink), some care is needed in controlling excited-state contamination so that masses are not found to be lower than their asymptotic value. Empirically, we have found the best way to do this is to form a simple $2\times 2$ generalised Pencil of Functions (PoF) \cite{Aubin:2010jc,Green:2014xba,Fischer:2020bgv} matrix of the Wall-Point correlator $C_\Omega^{\pm}(t)$
\begin{equation}
  M_\Omega^{\pm}(t) = 
  \begin{pmatrix}
    C_\Omega^{\pm}(t) & C_\Omega^{\pm}(t+1) \\
    C_\Omega^{\pm}(t+1) & C_\Omega^{\pm}(t+2) \\ 
  \end{pmatrix}.
\end{equation}
We solve this as a symmetric Generalised Eigenvalue Problem (GEVP) \cite{Michael:1982gb,Luscher:1990ck} at a fixed time $\tau_0$ and use the eigenvectors at some later time $\tau$ to approximately diagonalise the resulting correlator matrix. We then take the diagonal components as the principal correlators. A simple, fully-correlated, single-exponential fit to the lowest-lying principal correlator is sufficient to determine the ground state $\Omega$-baryon mass precisely. Some attention is needed in changing the GEVP parameters $\tau_0$ and $\tau$, and increasing these for finer lattice spacings is necessary. (Similar results could also be obtained by increasing the spacing between elements creating the matrix.) The PoF method yields a statistically-noisier result but removes a large amount of the excited-state contamination, which would usually cause the Wall-point correlator to approach from below. The application of the PoF is a trade-off as it allows for fits at short times albeit with data which is noisier than usual. The PoF also removes a lot of correlation within neighbouring points of the data making correlated fits easier. It also agrees well with a correlated two-exponential fit, with much less ambiguity on the ground-state's mass as compared to multi-exponential fits which often show some lower fit-bound dependence. A comparison of these two approaches can be seen in Fig.~\ref{fig:C101_pof} for one of our ensembles; note that there is no reduction in statistical uncertainty but the PoF is less sensitive to the choice of fit range.

Our final approach is quite similar to the scale-setting used in \cite{Borsanyi:2020mff}, but in that work the authors used smeared point sources, missing out on the volume average from the wall source. The PoF is not ubiquitously effective, for instance it does not often work well for mesons; in \cite{Hudspith:2023loy} we found it was effective for only one of our tetraquark operators, the one that looked like a compact diquark-antidiquark. This suggests that the method could have something to do with the spin-color nature of the operator and the compact nature of the state considered.

%% file: fitform.tex
\section{A flavor-$\text{SU}(3)$ chiral fit\label{chiral}}

We apply the chiral Lagrangian with three light flavours to compute the quark-mass dependence of the $\Omega$ mass. Expressions for all baryon octet and decuplet masses have been documented in a series of previous works \cite{Semke:2005sn,Lutz:2014oxa,Lutz:2018cqo,Guo:2019nyp,Lutz:2023xpi} where particular emphasis was put on the use of on-shell hadron masses in the one-loop contributions. In previous works \cite{Semke:2011ez,Lutz:2014oxa,Lutz:2018cqo} such a framework lead to significantly-improved convergence properties of the chiral expansion. Moreover, with the use of on-shell masses a much-improved description of finite-box effects stemming from such loop terms was seen \cite{Lutz:2014oxa}. In particular, the well-known L\"uscher type terms are contained in our expressions, as is important for most of the baryon decuplet states. 
Given such a framework in the isospin-limit, the set of baryon masses on a specified lattice ensemble would have to be determined as a solution of 8 coupled and non-linear equations.

In this work we focus on the positive-parity ground-state $\Omega$-mass on CLS ensembles, where we take the lattice values of all hadron masses that enter the loop expressions, i.e. the pion and kaon masses as well as $J=1/2$ and $3/2$ $\Xi$ masses. 
Instead of solving the aforementioned non-linear system we insert lattice values into the loop expressions and extract from there specific combinations of Low-Energy-Constants (LECs), and ultimately the lattice scale parameters. Our starting point is an expression for the $\Omega $-mass valid in a finite box at the N$^3$LO accuracy level. As will be illustrated below in Sec.~\ref{sec:n3lofit} an attempt fitting N$^2$LO 
does not suffice to arrive at the precision required for a faithful reproduction of the lattice data. We note that an extrapolation attempt using the chiral Lagrangian with two light flavours only \cite{Tiburzi:2008bk} is futile also in our case. 

In the following we denote with $m_\pi$, $m_K$ and $m_\eta$ the masses of the Goldstone bosons at a given lattice scale ``a" of a considered ensemble. Correspondingly, $M_\Omega$, $M_\Xi$ and $M_{\Xi^*}$  are the baryon masses on that ensemble. Later on we will also consider such masses as expressed in units of $8\,t_0$, a popular choice in the lattice community. We fit our data in terms of the dimensionless quantities on each lattice ensemble. Labeling these for a particular state ``Q" as,
\begin{equation}
\varphi_Q = 8t_0 m_Q^2\;.
\end{equation}
Our starting point will involve specific combinations of LECs used previously in \cite{Lutz:2023xpi}.

Any application of the chiral Lagrangian requires the  knowledge of the quark masses, $m_u , m_d,$ and $m_s$, that characterize a given lattice ensemble.  In principle, the $\pi$, $K$, and $\eta$ masses follow in terms of a set of relevant LECs. If the loop contributions to the meson masses are formulated in terms of on-shell masses, already the determination of the the $\pi$, $K$, and $\eta$ go in terms of the solution of a set of three coupled and nonlinear equations. Conversely, if $\pi$ and $K$ masses are known on a given ensemble, we may directly compute the quark masses \cite{Lutz:2018cqo,Lutz:2023xpi}. This is exemplified by
\begin{equation}\label{res-quark-mass-difference}
\begin{aligned}
B_0(m_s -m) &= m_K^2 -m_\pi^2 + \frac{m_\pi^2}{2\,f^2}\,\bar I^{(0)}_\pi - \frac{m_\eta^2}{2\,f^2}\,\bar I^{(0)}_\eta + \frac{m_K^2- m_\pi^2}{3\,f^2} \,\bar I^{(0)}_\eta \\
&-8\,\frac{m_K^4-m_\pi^4}{f^2}\,\big( 2\,L_8 -L_5\big) - 8\,\frac{m_K^2-m_\pi^2}{f^2}\,(2\,m_K^2+ m_\pi^2)\,\big(2\,L_6 -L_4 \big) \,, \\
B_0(m_s +2m) &=  m_K^2 +m_\pi^2/2 - \frac{m_\pi^2}{4\,f^2}\,\bar I_\pi^{(0)}  -   \frac{m_\eta^2}{4\,f^2} \,\bar I_\eta^{(0)} \\
&-8\,\frac{m_K^4+ m_\pi^4/2 }{f^2}\,\big( 2\,L_8 -L_5\big) 
-4\frac{2\,m_K^2+m_\pi^2}{f^2}\,(2\,m_K^2+ m_\pi^2)\,\big(2\,L_6 -L_4 \big) \,,
\end{aligned}
\end{equation}
where we used $m_\eta ^2= (4\,m_K^2-m_\pi^2)/3$ in terms that constitute corrections to the Gell-Mann--Okaes--Renner (GMOR) relations \cite{Gell-Mann:1968hlm} and assume an isospin-symmetric world with $m=m_u=m_d$ for simplicity. The LECs $B_0$ and $L_{4,6,7,8}$ are from Gasser and Leutwyler \cite{Gasser:1983yg}. The finite-box tadpole functions $\bar I_{\pi, K, \eta}^{(0)}$ are specified in Appendix \ref{app:chiral_extra}.

With this preparation we can write down an extrapolation form for the $\Omega$ mass only in terms of on-shell masses. We find
\begin{equation}\label{res-MOmega}
\begin{aligned}
M_\Omega &= M+ \Delta -4\,(d_0+ d_D/3)\,(m_K^2 +m_\pi^2/2) -\frac{8}{3}\,d_D \,(m_K^2- m_\pi^2 ) \\
&+\, \frac{1}{f^2}\,C_A^2\, J_{K \Xi }(M_\Omega)/Z_\Omega+\,  \frac{1}{3\,f^2}\,H_A^2\, J_{\eta \Omega}/Z_\Omega  +\, \frac{1}{3\,f^2}\,H_A^2\, J_{K \Xi^*}(M_\Omega )/Z_\Omega\\
&- \frac{1}{f^2}\,\sum_{Q =\pi, K, \eta}\Big( g_{ \Omega Q}^{(S)} \,m_Q^2\,\bar I^{(0)}_Q + g_{\Omega Q}^{(V)}\,\bar I_Q^{(2)}\Big)+ e^{(\pi)}_\Omega \,m_\pi^4 + e^{(K)}_\Omega \,m_K^4  \\
& +\, e^{(\eta )}_\Omega \,(m_K^2 -m_\pi^2) \,(4\,m_K^2 - m_\pi^2)/3\,
\end{aligned}
\end{equation}
with,
\begin{equation}
Z_\Omega = 1 + \frac{1}{f^2}\,\frac{d}{d \,M_\Omega} \Big\{ C_A^2\, J_{K \Xi }(M_\Omega) +  \frac{1}{3}\,H_A^2\,\Big[ J_{\eta \Omega}(M_\Omega) + J_{K \Xi^*}(M_\Omega ) \Big] \Big\},
\end{equation}
where $M+\Delta$ is the chiral-limit mass of the Omega. This requires the use of on-shell $\pi, K,\eta$ and $\Xi, \Xi^*, \Omega $ masses.  
Various combinations of LECs are encountered with $d_0,d_D, C_A, H_A$ and $g^{(S,V)}_{\Omega Q}$, $e_\Omega^{(Q)}$ with $Q =\pi, K,\eta$. For instance, at tree-level one may have a first rough estimate
\begin{eqnarray}\label{Eq:dd_tree}
 d_D \simeq - \frac{3}{4} \,\frac{M_\Omega - M_{\Xi^*} }{m_K^2- m_\pi^2} \simeq -0.48 \,{\rm GeV} ^{-1}\,.
\end{eqnarray}
A simple estimate is possible  also for $ C_A/ f$ with $f\simeq 92$ MeV and $C_A \simeq 1.6 $ from the hadronic decay widths of the baryon decuplet states (see e.g. \cite{Semke:2005sn}). Meaningful estimates for the remaining parameters are much more challenging and typically require input from lattice QCD simulations \cite{Guo:2019nyp,Lutz:2023xpi}. The functions used here are detailed in App.~\ref{app:chiral_extra}.

While the particular form (\ref{res-MOmega}) is quite convenient in the context of a chiral extrapolation, it is not very appealing in the lattice QCD context. Therefore we use a rewrite in terms of quantities made dimensionless by means of the lattice $8\frac{t_0}{a^2}$. We have
\begin{equation}\label{res-MOmega-Lattice}
\begin{aligned}
\frac{ aM_\Omega^{\rm Latt.} }{M^{\rm Phys.}_\Omega} &= \;a\,\Big\{ 1 
-4\,(\tilde d_0 +\tilde d_D/3)\Delta(\tilde m_K^2 + \tilde m_\pi^2/2 )
-\frac{8}{3}\,\tilde d_D \,\Delta [\tilde m_K^2- \tilde m_\pi^2 ] 
 \\
& +\, \frac{1}{\tilde f^2}\, \tilde c_\Omega^2\, \Delta [\tilde J_{K \Xi }( \tilde M_\Omega)/Z_\Omega ]
 +\,  \frac{1}{3\,\tilde f^2}\,\tilde h_\Omega^2\,\Delta [\tilde J_{\eta \Omega}(\tilde M_\Omega)/Z_\Omega  ]
 + \frac{1}{3\,\tilde f^2}\,\tilde h_\Omega^2\,\Delta [ \tilde J_{K \Xi^*}(\tilde M_\Omega ) /Z_\Omega ]
\\
& 
-\, \frac{1}{\tilde f^2}\,\sum_{Q =\pi, K, \eta}\Big( \tilde g_{ \Omega Q}^{(S)} \,\Delta [\tilde m_Q^2\,\tilde  I^{(0)}_Q  ]+ \tilde g_{\Omega Q}^{(V)}\,\Delta[ \tilde I_Q^{(2)} ]\Big)
 \\
& +\, \tilde e^{(\pi)}_\Omega \,\Delta [\tilde m_\pi^4 ]+ \tilde e^{(K)}_\Omega \,\Delta [\tilde m_K^4 ] 
+ \tilde e^{(\eta )}_\Omega \,\Delta [ (\tilde m_K^2 -\tilde m_\pi^2) \,(4\,\tilde m_K^2 - \tilde m_\pi^2)/3 ]\Big\}\, ,
\end{aligned}
\end{equation}
with the following conversions
\begin{equation}\label{res-MOmega-Lattice_pars}
\begin{aligned}
& \varphi_Q=   \tilde m_Q^2 = 8\,t_0\,m_Q^2 \,,\quad \tilde d_{0/D}  = \frac{1}{8t_0 M_\Omega^{\rm Phys.}}\, d_{0/D}  \,,\quad \;\;
 \tilde e^{(Q)}_\Omega  = \frac{1}{(8 \,t_0)^2\,M_\Omega^{\rm Phys.}}\, e^{(Q)}_\Omega  \,,\\
& \tilde c_\Omega/\tilde{h}_\Omega  =\sqrt{ \frac{1}{\sqrt{8 \,t_0} M_\Omega^{\rm Phys.}} }\,C_A/H_A \,,\quad\quad
\tilde g^{(S/V)}_{\Omega Q}  =  \frac{1}{8 \,t_0\,M_\Omega^{\rm Phys.}}\, g^{(S/V)}_{\Omega Q}\,\\
& Z_\Omega =1+  \frac{ \tilde M^{\rm Phys.}_\Omega }{\tilde f^2}\,\frac{d}{d\,\tilde M_\Omega} \,\Big\{ 
 \tilde c_\Omega^2 \,\tilde J_{K \Xi} (\tilde M_\Omega ) 
 + \frac{1}{3}\,\tilde h_\Omega^2\,\Big( \tilde J_{\eta \Omega}(\tilde \Omega)+\tilde J_{K \Xi^*}( \tilde M_\Omega) \Big)\Big\}\,.
\end{aligned}
\end{equation}
They are obtained from the functions in (\ref{res-MOmega}) by a multiplication with suitable powers of $8t_0$. Here the masses and $t_0$ values as measured on each ensemble are used. In Eq.~\ref{res-MOmega-Lattice} we use the notation $\Delta [\cdots ] $ to indicate that the argument is subtracted by its corresponding expression at the physical point, i.e. at physical quark masses, infinite volume, and in the continuum limit. By making this subtraction point a fit parameter we can provide a measurement for the continuum $t_0$.

From a practical perspective this formulation is neat because everything within the curly braces in Eq.~\ref{res-MOmega-Lattice_pars} is dimensionless and can be fit simultaneously over all ensembles, dividing out the lattice-spacing dependence. Comparing this to Eq.~\ref{res-MOmega-Lattice}, which requires everything (LECs and all) to have an implicit scale, our lattice fit omits this tedious book-keeping exercise. $t_0$ on each ensemble is often more precisely determined than the $K$ or $\pi$ so forming these dimensionless quantities does not incur any difficulties.

Eqs.~\ref{res-MOmega} and \ref{res-MOmega-Lattice_pars} are organised by order in the chiral expansion. With the ``1" being the leading order (LO), the terms $d_0$ and $d_D$ contributing at next-to-leading-order (NLO), the ``bubble" terms $C_A$ and $H_A$ at N$^2$LO and the final ``tadpole" terms $g_{\Omega(\pi/K/\eta)}^{(S/V)},e_\Omega^{(\pi/K/\eta)}$ at N$^3$LO. For Eq.~\ref{res-MOmega-Lattice_pars} the NLO terms will also receive corrections from the small mass-dependence of $t_0$ at N$^3$LO, which we will effectively parametrize.

Altogether there are 13 combinations of LECs involved, with terms $d_0,\, d_D $ and $ C_A,\, H_A$ and  $ e^{(Q)}_\Omega,\, g^{(S)}_{\Omega Q},\, g_{ \Omega Q}^{(V)}$. Most important are $d_0, d_D\,,C_A$, $g^{(S)}_{K \Omega}+g^{(S)}_{\eta  \Omega}$,\, $e^{(\eta)}_\Omega$, which already permit an excellent reproduction of the data set. Our current dataset is not able to determine all LECs that are relevant for the $\Omega$-mass at 
N$^3$LO. This is expected to change once further ensembles are considered.

%% file: results.tex
\section{Lattice determination of the Omega-baryon masses}\label{sec:latres1}

In Tab.~\ref{tab:omegas} we list our results (in lattice units) for the $I(J^P)=0(3/2^+)$ and $0(3/2^-)$ $\Omega$-baryon masses as well as the statistics used to obtain our measurement. Further details on the generation of these $n_f=2+1$ Wilson-clover ensembles and the naming convention used can be found in \cite{Bruno:2014jqa}. It is evident that as the strange mass increases to the physical point and as the physical volume increases, the statistical error of the determined $\Omega$-baryons decreases significantly for the positive-parity state but for the negative-parity the situation is less clear.

\begin{table}[p]
\footnotesize
\begin{tabular}{c|ccc|cc}
  \toprule
  Ensemble & T & $L^3\times T$ & $N_\text{S}\times N_\text{C}$ & $aM_\Omega^{(3/2)^+}$ & $aM_\Omega^{(3/2)^-}$ \\
  \hline
  A653 & Periodic & $24^3\times 48$ & $48\times 1000$ & 0.7045(16) & 0.8833(28) \\
  A654 & Periodic & $24^3\times 48$ & $48\times 1000$ & 0.7439(12) & 0.9142(21) \\
  B650 & Periodic & $32^3\times 64$ & $64\times 166$  & 0.7558(16) & 0.9287(24) \\
  \hline
  U103 & Open     &  $24^3\times 128$ & $68\times 132$  & 0.6334(24) & 0.7915(27) \\
  H101 & Open     & $32^3\times 96$ & $48\times 500$  & 0.6169(12) & 0.7755(37) \\
  H102 & Open     & $32^3\times 96$  & $48\times 125$  & 0.6422(13) & 0.7985(10) \\
  H105 & Open     & $32^3\times 96$  & $48\times 125$  & 0.6681(14) & 0.8245(21) \\
  N101 & Open     & $48^3\times 128$ & $96\times 100$  & 0.6631(13) & 0.8186(19) \\
  C101 & Open     & $48^3\times 96$  & $48 \times 126$ & 0.6794(14) & 0.8309(18) \\
  \hline
  B450 & Periodic & $32^3\times 64$  & $64\times 200$   & 0.5549(28) & 0.6974(24) \\
  S400 & Open     & $32^3\times 128$ & $64\times 457$   & 0.5788(26) & 0.7171(8) \\
  X451 & Periodic & $40^3\times 128$ & $64\times 500$   & 0.5936(7)  & 0.7293(6) \\
  N451 & Periodic & $48^3\times 128$ & $128 \times 200$ & 0.5946(7)  & 0.7316(8) \\
  D450 & Periodic & $64^3\times 128$ & $128\times 125$  & 0.6111(7)  & 0.7491(15) \\
  D452 & Periodic & $64^3\times 128$ & $128\times 96$   & 0.6164(8)  & 0.7550(14) \\
  \hline
  H200 & Open     & $32^3\times 96$  & $48 \times 125$ & 0.4753(30) & 0.5933(41) \\
  N202 & Open     & $48^3\times 128$ & $64 \times 175$ & 0.4649(21) & 0.5859(23) \\
  N203 & Open     & $48^3\times 128$ & $64\times 190$  & 0.4882(11) & 0.6092(11) \\
  N200 & Open     & $48^3\times 128$ & $64\times 102$  & 0.5068(14) & 0.6229(5) \\
  D200 & Open     & $64^3\times 128$ & $64\times 95$   & 0.5218(6) & 0.6352(8) \\
  E250 & Periodic & $96^3\times 192$ & $192\times 62$  & 0.5285(3) & 0.6391(35) \\
  \hline
  N300 & Open     & $48^3\times 128$ & $64\times 192$ & 0.3676(23) & 0.4578(35) \\
  N302 & Open     & $48^3\times 128$ & $64\times 100$ & 0.3871(15) & 0.4799(18) \\
  J303 & Open     & $64^3\times 192$ & $64\times 63$  & 0.4023(14) & 0.4906(8) \\
  E300 & Open     & $96^3\times 192$ & $120\times 63$ & 0.4138(6)  & 0.4959(12) \\
  \hline
  J500 & Open     & $64^3\times 192$ & $72 \times 94$ & 0.2851(18) & 0.3618(16) \\
  J501 & Open     & $64^3\times 192$ & $96\times 62$  & 0.2996(8)  & 0.3765(16) \\
  \botrule
\end{tabular}
\caption{Omega masses calculated for the determination of the lattice spacing. $N_S$ indicates the number of propagator solves per gauge configuration, and $N_C$ the number of independent gauge configurations used.}\label{tab:omegas}
\end{table}

In this work we use a subset of generated configurations for our measurement, choosing them maximally-evenly-spaced throughout the Monte-Carlo trajectory. Making use of the improved deflated reweighting factors from \cite{KUBERSKI2024109173} (we happen to use sets of configurations where there is mostly no evidence for negative strange reweighting factors \cite{Mohler:2020txx}, exceptions were the ensembles A654, S400,  D452, D450, and E250). Once this is done, some binning is performed when necessary in order to reduce autocorrelations. As we do not have the raw bootstraps for the $\varphi_Q$-variables, random Gaussian noise  distributions are drawn with widths corresponding to the error. Not knowing the correlation between these measurements we treat them as fully uncorrelated. As they are ``x-axis errors" we don't expect this to lead to a significant uncertainty in determining the lattice spacings accurately (the slope in mass turns out to be small) but at close to physical masses the relative uncertainty of the pion mass in particular will have a significant contribution to the total uncertainty of the final determination of the lattice scale.

\begin{figure}[h]
  \includegraphics[scale=0.55]{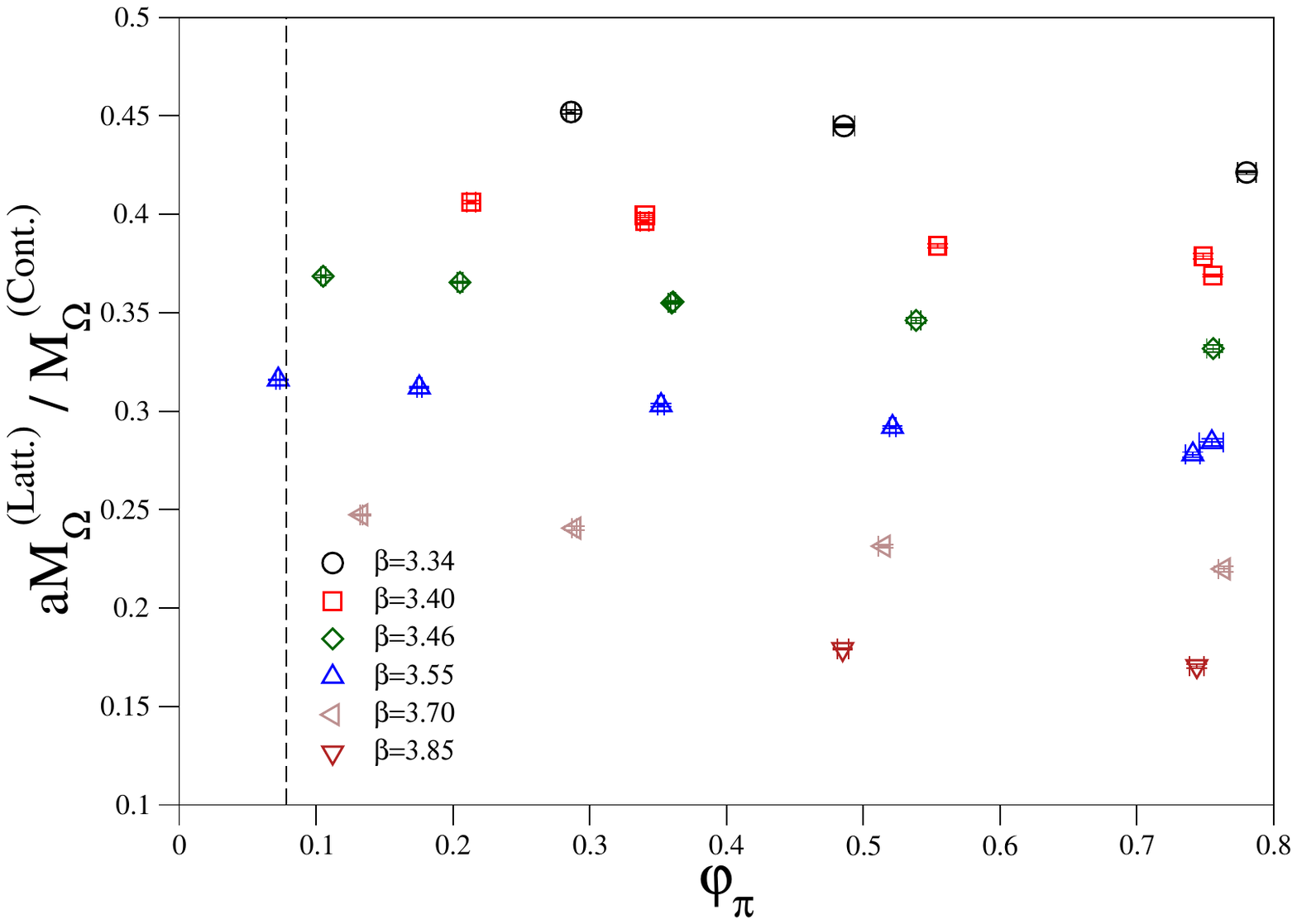}
  \caption{Our ($aM_\Omega^{(3/2)^+}$) data (divided by the continuum $\Omega$-baryon in GeV units) vs $\varphi_\pi$. The dashed line represents the $n_f=2+1$ iso-symmetric continuum value of $\varphi_\pi$ using our fitted $t_0$.}\label{fig:lattdata}
\end{figure}

Several of these ensembles were generated with the same action parameters, differing only in volume. (U103, H101) and (H200, N202) are at the $\text{SU}(3)_f$-symmetric point, whereas (H105, N101) and (X451, N451) are at intermediate pion mass. Having these pairs of ensembles provides a very strong handle on finite-volume effects. Already we can see from Tab.~\ref{tab:omegas} and Fig.~\ref{fig:lattdata} that the largest finite-volume effects for our ensembles occur where the $\pi$, $K$, and $\eta$ are degenerate. Indeed, U103 has $m_{\pi/K/\eta} L$ of $4.33$ whereas H101 has $5.83$, similarly H200 has $4.33$ and N202 has $6.44$.

At intermediate pion mass there are large $m_\pi L$ differences between H105 and N101 ($3.88$ and $5.82$) yet tiny differences in their $\Omega$-baryon masses. Along with the $\text{SU}(3)_f$-symmetric point data this strongly indicates finite volume effects are dominated by $m_K L$ or $m_\eta L$ ($6.47$ and $9.68$ for $m_K L$ on these boxes). Similarly, X451 and N451 have $m_\pi L=(4.42,5.31)$ and $m_K L=(7.12,8.56)$ respectively, but yet they have statistically compatible results for $aM_\Omega^{(3/2)^+}$. These few ensembles already provide strong-enough handles in $m_\pi L$ to rule-out a pure $\text{SU}(2)_f$ chiral description, as that formally has no $m_K L$ contribution.

Fig.\ref{fig:lattdata} illustrates our lattice data divided by the PDG $\Omega$-baryon mass ($1.67245$ GeV). From the coarsest (black) to the finest (indigo) ensembles, against the dimensionless pion-mass-squared analog $\varphi_\pi$.

\subsection{Comparison of $aM_\Omega^{(3/2)^+}$ with previous determinations}

\begin{figure}[tb]
\includegraphics[scale=0.55]{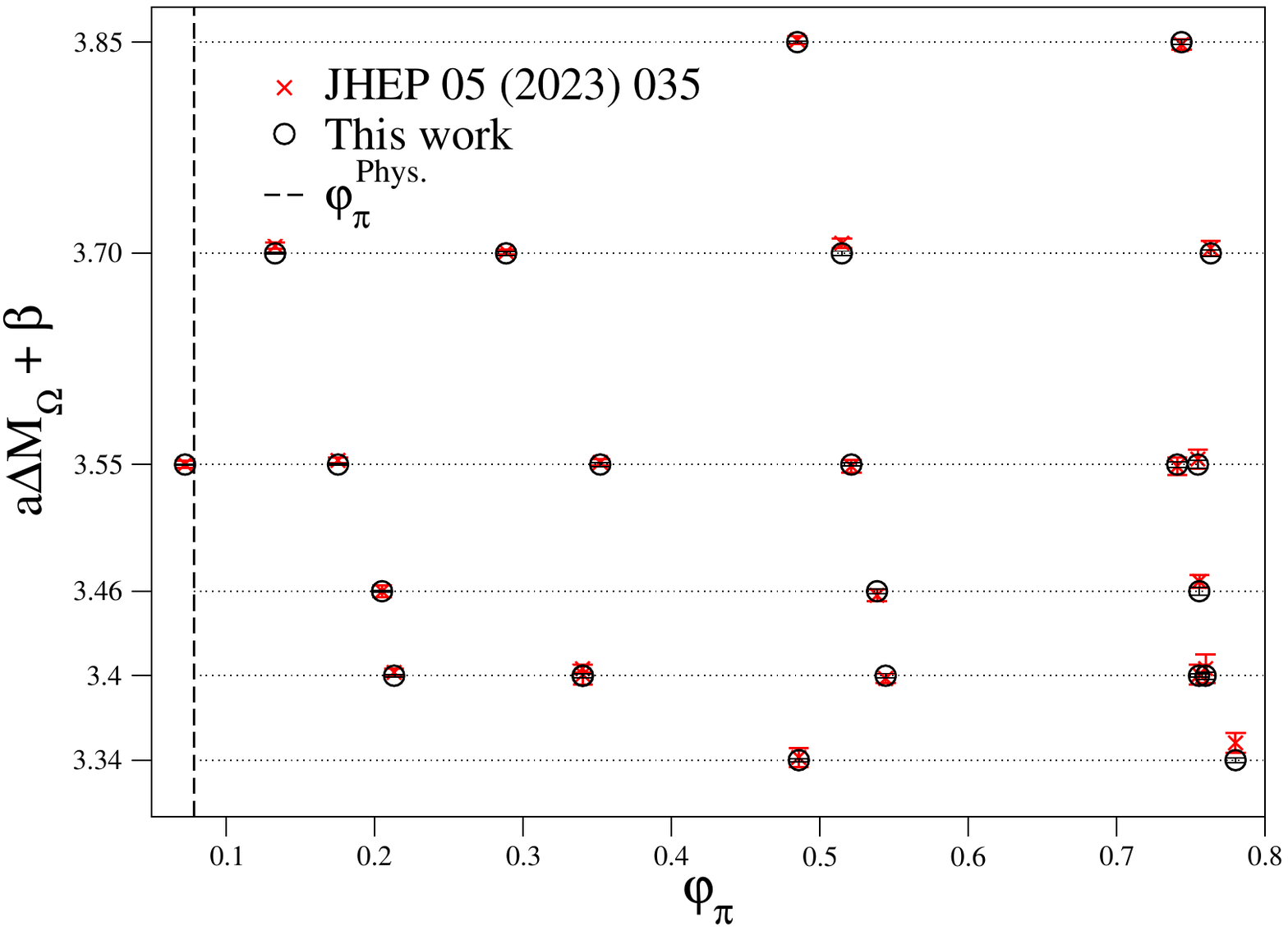}
\caption{A comparison of our bare $aM_\Omega^{(3/2)^+}$ results with those of \cite{RQCD:2022xux} upon subtracting our central value and adding $\beta$, excellent consistency is seen across all ensembles shared by the two analyses.}\label{fig:regbeta}
\end{figure}

\begin{figure}[h]
\includegraphics[scale=0.55]{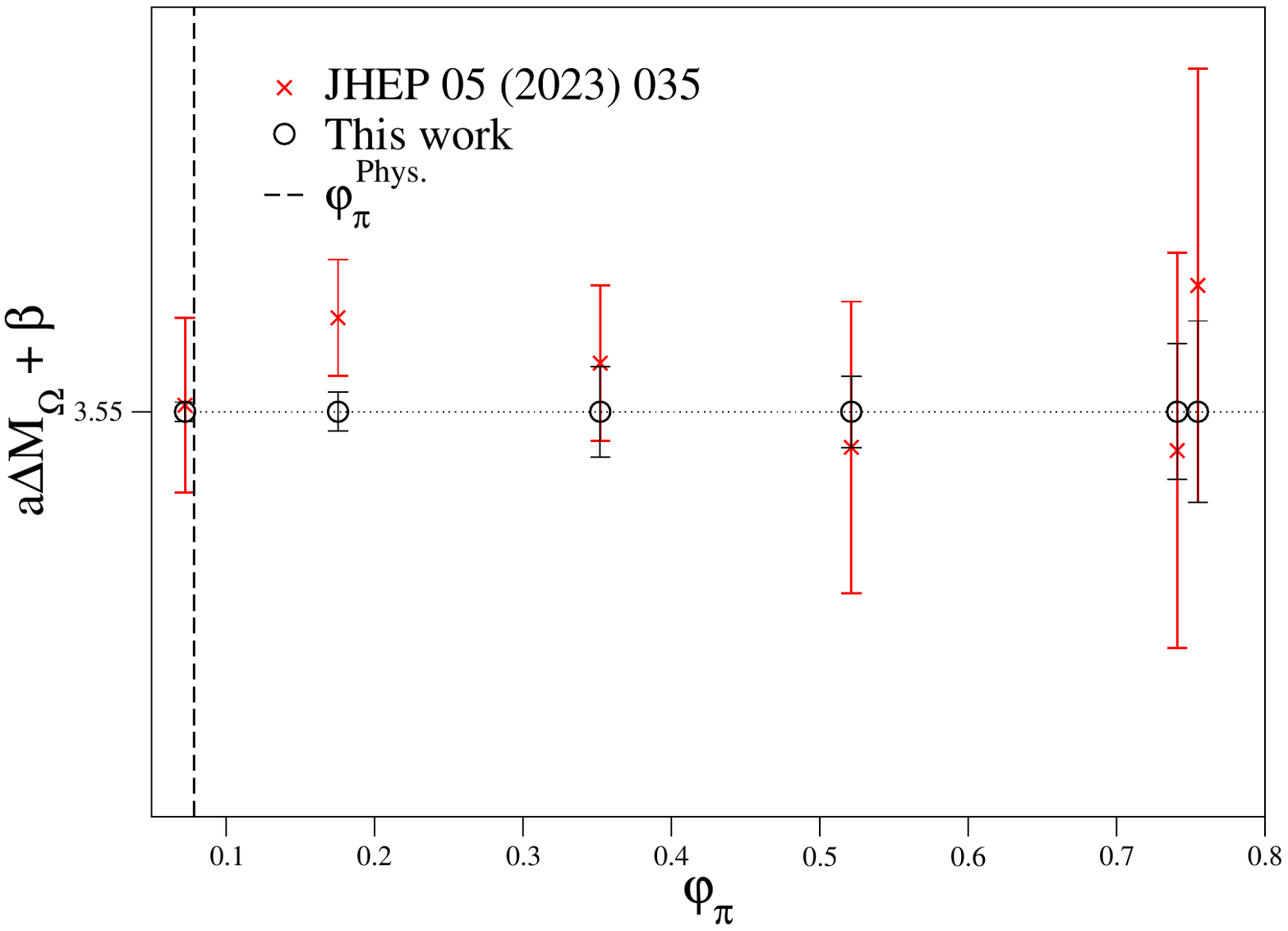}
\caption{Same as Fig.\ref{fig:regbeta} but with a zoom into the data for $\beta=3.55$, illustrating the improvement in the statistical uncertainty of our mass-determination and the quantitative agreement of the results with \cite{RQCD:2022xux}.}
\label{fig:2latzoom}
\end{figure}

The authors of \cite{RQCD:2022xux} have measured independently the $\Omega$-baryon on most of the  ensembles used here, so it is important to cross-check with their results. Fig.~\ref{fig:regbeta} shows the bare (lattice units) mass determinations where we have subtracted our central value from both results (where results exist on the same ensembles) and added the bare gauge coupling $\beta$ to distinguish data points in a plot. This is plotted again against $\varphi_\pi$ to check for consistency as a function of the light-quark mass. Our data is at least a factor of 2 statistically more precise on all ensembles, and at best has a factor of 10 smaller statistical uncertainty. Our largest discrepancies with \cite{RQCD:2022xux} are for A653 $(1.7\sigma)$, D200 $(1.6\sigma)$, and E300 $(2.3\sigma)$. In cases where there exists discrepancies their results are typically systematically above ours. As the authors of \cite{RQCD:2022xux} use Smeared-Smeared point sources, which are bounded from above, this could indicate that there is still some excited-state contamination in their mass determinations.

Figure~\ref{fig:2latzoom} shows a zoom into the $\beta=3.55$ data of Fig.~\ref{fig:regbeta}. Here we see that our statistical resolution is always better and dramatically improves as the physical volume increases, due to the better volume average obtained with gauge-fixed wall sources. This is an important feature for our mass trajectory: As our closest-to-physical pion masses also have the largest extent, and as our lattice spacing is defined at physical masses these will provide the greatest constraint to our scale determination and $t_0$. In a cost comparison to \cite{RQCD:2022xux} for the largest box (i.e.~most computationally-expensive) with the lightest pion mass (E250) they used in total $15,648$ smeared, full precision, point sources. In comparison, we used $11,904$ mostly low-precision wall sources. Even for the strange quark our truncated solves are a factor of 4 cheaper than full solves. Comparatively, we are roughly a factor 5 cheaper in overall cost, and almost a factor 10 more statistically precise. One should however note that our calculation is a dedicated project while theirs is performed as part of a larger baryon-structure physics programme.

%% file: results_fits.tex
\section{Results of the chiral fits and determination of the lattice scale}\label{sec:latfits}

Our fits to the data in Tab.~\ref{tab:omegas} are all based on Eq.~\ref{res-MOmega-Lattice}, where a few remarks are in order. In the upper part of Tab.~\ref{tab:finalfit_fermi} we list the chiral input parameters that we take from the literature \cite{Lutz:2023xpi}. Anticipating that our Omega-baryon fits are not discriminative to these choices, these parameters can at least be partially absorbed into our fit parameters. Already at N$^3$LO we have too many fit parameters and will not be able to obtain values for these ourselves. The pole masses and $t_0$ values in Eq.~\ref{res-MOmega-Lattice} are mostly taken from \cite{Ce:2022kxy} and \cite{RQCD:2022xux} with the exception of those directly measured by us and listed in App.~\ref{app:further_measurements}.

The $\eta$-contributions are estimated  using the GMOR relation with  $m_\eta ^2= (4\,m_K^2-m_\pi^2)/3$. It turns out that for the fit the finite-volume tadpole terms $\tilde I_\eta^{(n)}$ are quite similar to those of $\tilde I_K^{(n)}$ in (\ref{res-MOmega}). Along our trajectory it is impossible to separate the two types of contributions. Keeping this in mind we drop all tadpole and bubble terms involving the $\eta$, assuming that their impact is taken care of by using effective kaon parameters. Unfortunately, with the precision of the current data we must add a few priors for chiral parameters for some of the flattest directions, these being: $\tilde{d}_D=-0.07\pm 0.03$, $\tilde{c}_\Omega=0.9\pm 0.9$, and $\tilde{h}_\Omega=0.4\pm 0.2$. Ultimately the two bubble terms do not contribute significantly, but due to the dependence of $\tilde{Z}_\Omega$ on $\tilde{c}_\Omega$ and $\tilde{h}_\Omega$ we need something to help stop the fitter wandering off. The prior on $\tilde{d}_D$ is based on Eq.~\ref{Eq:dd_tree}. An argurment for the prior on $\tilde{c}_\Omega$ has been made earlier, whereas the prior on $\tilde{h}_\Omega$, which is poorly known, is from an ad-hoc procedure. We set it by performing a fit with a wide prior and moving the central value to the former fit result and shrinking the width until it is $50\%$ width of the central value. The fit struggled with the wide prior and this aforementioned procedure was found to most-stably fit the data. Regardless of this procedure the lattice spacings remain the same, even after setting $\tilde{h}_\Omega$ to zero.

For the iso-symmetric kaon and pion that define our physical light-mass point we use the values recommended in \cite{Aoki:2016frl} of $m_\pi=134.8(3) \text{ MeV}$ and $m_K=494.2(3) \text{ MeV}$. In Eq.~\ref{res-MOmega-Lattice} in the J-functions, the $\Xi,\,\Xi^*$ enter; for these we use the measured lattice masses but we must subtract their continuum values in the definition of our $\Delta$. To do so, we take the masses from the 2023 PDG \cite{Workman:2022ynf}, forming an average of the (real part) of the charged and neutral states. For the physical $\Omega$ we will perform a fit using the PDG value and one using an estimate of the mass-correction due to QED, quoting the latter for our final results.

\subsection{Fit parameters - the need for N$^3$LO}\label{sec:n3lofit}

Table~\ref{tab:chi_order} indicates the quality of fit we obtain with increasing order in the expansion of Eq.~\ref{res-MOmega-Lattice_pars}. It should be clear from Fig.\ref{fig:lattdata} that we cannot fit this data sensibly to a constant, hence the LO value has a huge $\chi^2/dof$. Going to NLO helps significantly in providing a slope to the data but still is nowhere near a good fit. Adding the bubble terms provides only a mild improvement and adding the tadpole terms that come with the strong finite-volume effects improves the situation greatly. Here we drop the $\tilde{g}^{(S/V)}_{\Omega \eta}$ terms as the fit cannot distinguish them from the kaon contributions. We also set to zero $\tilde{e}_\Omega^{(\pi)}$ and $\tilde{e}_\Omega^{(K)}$ as they turn out to be zero, but keep $\tilde{e}_\Omega^{(\eta)}$ with the intention of absorbing any mass-dependence from $t_0$ \cite{Bar:2013ora} that would enter at N$^3$LO in our expressions.

\begin{table}[tb]
\begin{tabular}{@{\hspace{1em}}c@{\hspace{1em}}|@{\hspace{1em}}c@{\hspace{1em}}|@{\hspace{1em}}c@{\hspace{1em}}|@{\hspace{1em}}c@{\hspace{1em}}|@{\hspace{1em}}c@{\hspace{1em}}}
\toprule
LO & NLO & N$^2$LO & N$^3$LO \big(no $\tilde{g}_{\Omega\eta}^{(S/V}$\big) & N$^3$LO \big(no $\tilde{g}_{\Omega(\pi/\eta)}^{(S/V)}$\big)\\
\hline
323 & 4.1 & 3.1 & 0.54 & 0.69 \\
\botrule
\end{tabular}
\caption{$\chi^2/dof$ for our fit with increasing order of $\text{SU}(3)_f$ chiral expansion.}\label{tab:chi_order}
\end{table}

Already at N$^3$LO we are close to over-determining our data $p > 0.9$), and it turns out that we can reasonably omit the $\tilde{g}^{(S/V)}_{\Omega \pi/\eta}$ terms in Eq.~\ref{res-MOmega-Lattice} as $\tilde{g}_{\Omega\pi}^{(S)}$ was found to be consistent with 0 and $\tilde{g}_{\Omega\pi}^{(V)}$ only $1\sigma$ away from 0. This gives a slightly larger $\chi^2/dof$ with (27 data points, 14 fit parameters with 3 priors) 16 degrees of freedom and will be our final quoted result. Removing or keeping the pion tadpole terms in the fit does not change the obtained lattice spacings within error.

\subsection{Our final results for the scale}

Table~\ref{tab:finalfit_fermi} gives our final fit results based on Eq.~\ref{res-MOmega-Lattice} but without the extraneous parameters $\tilde{g}_{\Omega(\pi/\eta)}^{(S/V)}$, $\tilde{e}_\Omega^{(\pi)}$ 
and $\tilde{e}_\Omega^{(K)}$, which turns out to be zero with the renormalization scale we use. This parameterizes our preferred description of the $\Omega^{(3/2)^+}$ data in Tab.~\ref{tab:omegas}. The lattice (tilded) variables obtained by the fit are listed in Appendix~\ref{app:lattice_fits}. We convert our lattice spacings into fermi and our dimensionful fit parameters into GeV units using our determined value of $\sqrt{t_0}$ for self-consistency. The values are collected in Tab.~\ref{tab:finalfit} for final presentation and to compare with the literature.

The $\Omega$-baryon carries electric charge in nature, and this is expected to provide a correction to its mass at below the $0.2\%$ level \cite{CSSM:2019jmq}. To estimate the influence of QED we have taken the prediction from \cite{RQCD:2022xux} of $M_\Omega^{(3/2)^+}=1.6695 \text{ GeV}$, which is close to this upper bound, and rerun our fit. We quote our final result as these values and a difference between using the charged $\Omega$ and this result as an associated systematic. For all cases the difference in lattice spacings and $t_0$ obtained are negligible compared to our statistical uncertainty as the physical matching point (x-axis) for the $\Omega$ also changes slightly and mostly compensates the small change in the y-axis.

\begin{table}[tb]
\begin{tabular}{@{\hspace{1em}}c@{\hspace{1em}}|@{\hspace{1em}}c@{\hspace{1em}}||@{\hspace{1em}}c@{\hspace{1em}}|@{\hspace{1em}}c@{\hspace{1em}}}
\toprule
$f$ [MeV] & 92.4 &$\mu$ [MeV]& 770\\
M [MeV] &804.3& $\Delta$ [MeV]& 1115.2 \\
\hline\hline
$a(\beta=3.34)$ & $0.09329(27)(5) \text{ fm}$ & $a(\beta=3.40)$ & $0.08262(18)(4) \text{ fm}$ \\
$a(\beta=3.46)$ & $0.07380(14)(4) \text{ fm}$ & $a(\beta=3.55)$ & $0.06268(10)(3) \text{ fm}$ \\
$a(\beta=3.70)$ & $0.04884(13)(3) \text{ fm}$ & $a(\beta=3.85)$ & $0.03806(12)(2) \text{ fm}$ \\
\hline
$d_0$ & $-0.39(13) \text{ GeV}^{-1}$ & $d_D$ & $-0.51(15) \text{ GeV}^{-1}$ \\
$C_A$ & $1.7(4)$ & $H_A$ & $0.6(2)$ \\
$g_{\Omega K}^{(S)}$ & $-13(4) \text{ GeV}^{-1}$ & $g_{\Omega K}^{(V)}$ & $48(12) \text{ GeV}^{-1}$ \\ 
$e_\Omega^{(\eta)}$ & $-0.13(4) \text{ GeV}^{-3}$ & $\sqrt{t_0}$ & $0.14480(32)(6) \text{ fm}$ \\
\botrule
\end{tabular}
\caption{Chiral parameters used in the fits (upper part) and final lattice fit results for the LECs (lower part). The first uncertainty is statistical and the second an estimate from QED effects. For the LECs we drop the QED error as it is tiny compared to the statistical. All other parameters in Eq.~\ref{res-MOmega-Lattice} are set to zero.}\label{tab:finalfit_fermi}
\end{table}

We now turn our focus to the determination of the LECs. Given that we use an $\text{SU}(3)_f$-framework, an uncertainty of only around (2-3) MeV is certainly unexpected. We find it surprising that with an N$^3$LO chiral approach it is possible to reproduce the very accurate $\Omega$-masses across the entire range of pions and kaons used here. A previous study, \cite{Lutz:2023xpi}, estimated a significantly larger associated uncertainty of about (10-15) MeV on the baryon masses from the chiral-series truncation order. Nonetheless, our values for $d_0, d_D, C_A$ and $H_A$ appear reasonable, unsurprisingly as they are mostly set by priors. However, the somewhat unexpectedly-large size of the tadpole parameters $g^{(S/V)}_{\Omega K}$ in Tab. \ref{tab:finalfit_fermi} may be a consequence of using the chiral approach outside its domain of validity. A comparison of our LECs with the corresponding ``more natural" values of \cite{Lutz:2023xpi} shows large discrepancies, even though the analyses are based on similar ensembles. We note that setting either $g^{(V)}_{\Omega K}$ or $g^{(S)}_{\Omega K}$ to zero yields a $\chi^2/dof \approx 1.2$ with the nonzero parameter becoming larger to accommodate this.

\begin{figure}[tb]
\includegraphics[scale=0.554]{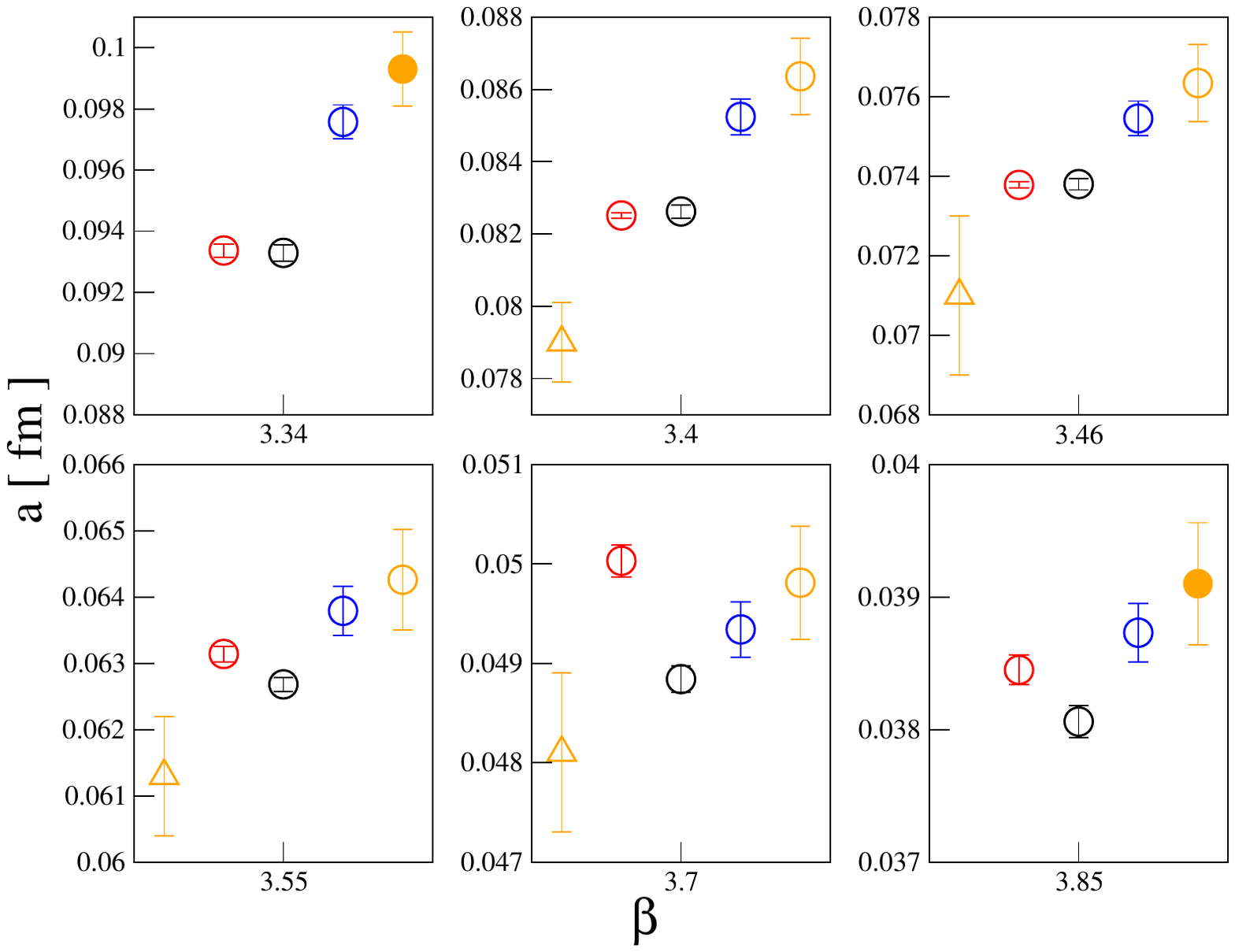}
\caption{Comparison of our lattice-spacing determinations (black circles, centre) with those in the literature for these ensembles (x-axis values have been shifted for clarity): ``Strategy 2" of \cite{Bruno:2016plf} (orange triangles, leftmost), \cite{Lutz:2023xpi} (red circles, first left),  \cite{RQCD:2022xux} (blue circles, first right), ``Strategy 1" of \cite{Bruno:2016plf} (orange circles, rightmost), and estimates from \cite{Chao:2022xzg} (filled circles) based on ``Strategy 1" of \cite{Bruno:2016plf} and $t_0$ at the $\text{SU}(3)_f$-symmetric point.}\label{fig:final_comp}
\end{figure}

The previous study of \cite{Lutz:2023xpi} used the data of \cite{RQCD:2022xux}, and draws into question some of the systematic uncertainties quoted in their obtained masses. Regardless of the fact that the baryon masses used in \cite{Lutz:2023xpi} were much less precise, the quoted precision in the lattice scales there came in part from a scale-setting in which the empirical values of all baryon masses were used. The quoted precision of that determination can be seen in the comparison with our results and others from the literature in Fig.~\ref{fig:final_comp}. While the true value of the chiral LECs remains a somewhat open question; in the future, improved lattice spectra should allow for a better determination of them. If our current values hold they may imply exotic features in the s-wave scattering processes of the Goldstone bosons off the baryon decuplet states. Whether or not the LECs are realistic is mostly irrelevant to the determination of the scale and $\sqrt{t_0}$. It is a form that describes our data well in interpolation/mild extrapolation (depending on the $\beta$) in the quark masses with an extrapolation in the volume, which is all we need.

As the physical fit parameter (and combinations with physical masses that incorporate it) $t_0$ is a constant, it competes with our determined lattice spacings (which are also constants for each $\beta$). It is therefore unsurprising that these parameters will be quite correlated (as can be seen by our correlation matrix in the Appendices in Tab.~\ref{tab:latcorr}). As the ensemble E250 ($\beta=3.55$) has the smallest statistical error and lies closest to the physical pion and kaon, it has the strongest influence on the fit, and yields the largest correlation with $t_0$. Our worst relative error ($0.32\%$) on the lattice spacing appears for $\beta=3.85$, where we only have two ensembles quite far from the physical pion and kaon masses. Our best relative error ($0.17\%$) is for the $\beta=3.55$ set of ensembles where we have many pion and kaon masses and a very precise determination below the physical quark-mass point, making this an interpolation in pion mass rather than an extrapolation.

\begin{figure}[tb]
\includegraphics[scale=0.4]{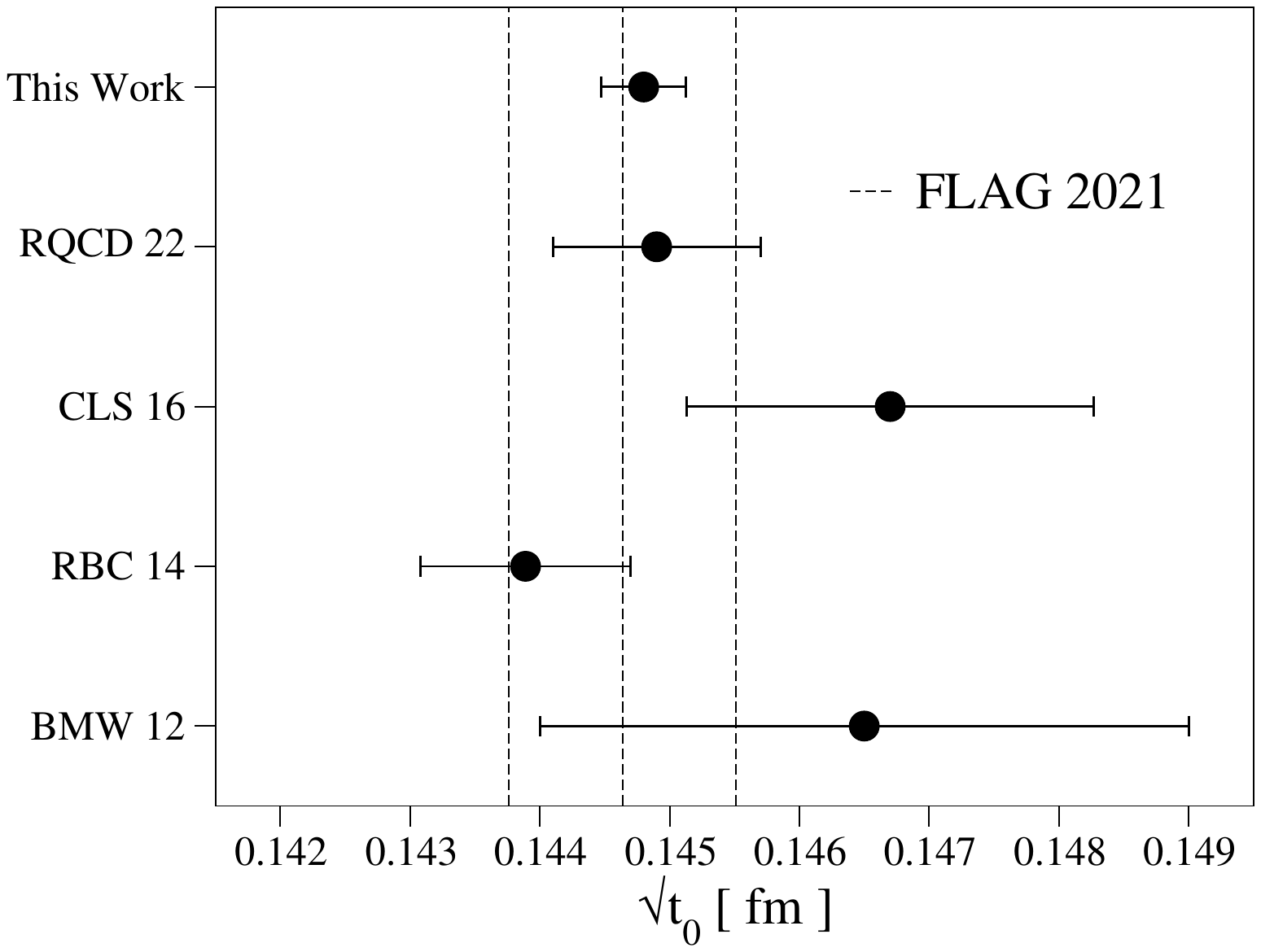}
\caption{A comparison of our result for $\sqrt{t_0}$ with the other, $n_f=2+1$, published determinations of \cite{RQCD:2022xux,Bruno:2016plf,RBC:2014ntl}, and \cite{BMW:2012hcm} plus the FLAG 2021 average \cite{FlavourLatticeAveragingGroupFLAG:2021npn} whose central value and error is drawn as a dashed line.}\label{fig:t0comp}
\end{figure}

Figure~\ref{fig:final_comp} shows our result compared to other determinations from this set of ensembles found in the literature. These should not necessarily agree at finite lattice spacing as they define a ``scheme" or ``renormalization trajectory" along which the chosen observables approach the continuum limit. They should however tend to agree at finer lattice spacing (larger $\beta$), and here we observe some tension with \cite{Lutz:2023xpi}. For this plot we have added the statistical and systematic uncertainties of all determinations in quadrature. It is interesting to note that our determinations are quite different to those from $f_{\pi K}$ \cite{Bruno:2016plf}, or $M_\Xi$ \cite{RQCD:2022xux}, and are more consistent with the determination of \cite{Lutz:2023xpi} based on baryon masses.

In Fig.~\ref{fig:t0comp} we compare our determination for the continuum $t_0$ for $n_f=2+1$, against others found in the literature as well as the last FLAG value \cite{FlavourLatticeAveragingGroupFLAG:2021npn}. Our result is consistent with all previous determinations, but with a significantly smaller total uncertainty due to the precision we were able to achieve in the measured masses of the $\Omega$-baryon. The top three determinations in the plot are not independent as they use some of the same configurations and good consistency is seen between them.

\section{The parity-odd state and the $\Omega(2012)^-$}\label{sec:32minus}

\begin{figure}[tb]
\includegraphics[scale=0.55]{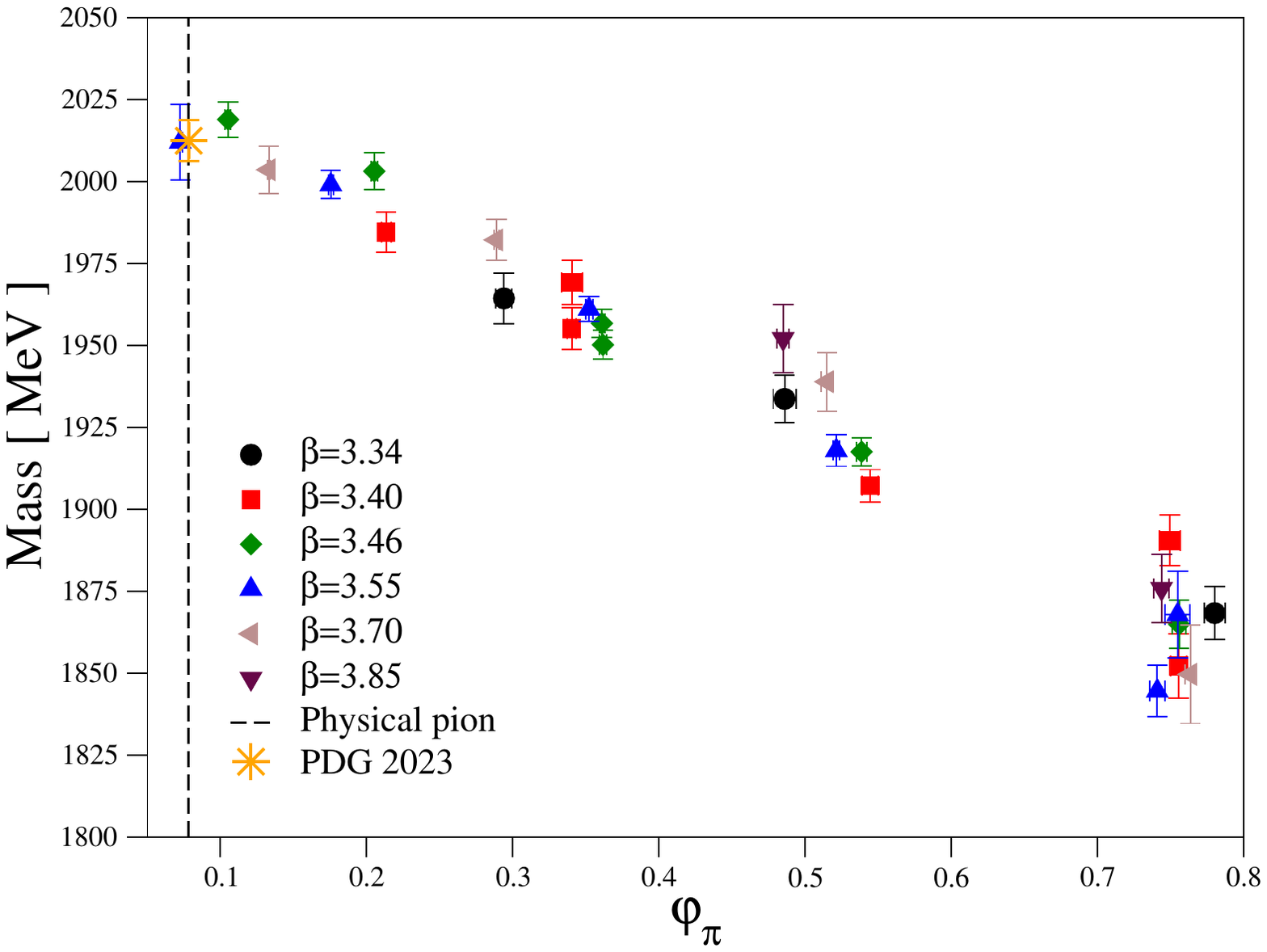}
\caption{Our negative-parity Omega masses (converted to MeV) compared to the PDG $***$-state \cite{Workman:2022ynf} $\Omega(2012)^-$, discovered by Belle \cite{PhysRevLett.121.052003}. here we have plotted the mass as the central value with the width of the state as the error. Note that along our trajectory the strange-quark becomes heavier as $\varphi_\pi$ approaches the physical value.}\label{fig:oddparity}
\end{figure}

In Tab.~\ref{tab:omegas} we listed both the parity-even and parity-odd $\Omega$ masses. Converting these results to physical units using the lattice-spacings in Tab.~\ref{tab:finalfit_fermi} gives the results for the negative-parity masses presented in Fig.~\ref{fig:oddparity}. Some care is warranted in interpreting these data; at intermediate pion masses this appears to be a compact object lying well below $\Xi^{*}K$, but at lower pion masses it approaches the (unstable) $\Xi^{*}K$, where the $\Xi^*$ decays strongly into $\Xi \pi$. This is likely why our uncertainty grows for smaller pion masses (and the ease of fitting decreases). Note that the experimental width of the  $\Omega(2012)^-$ is narrow as its decay is restricted only to the d-wave $\Xi K$ \cite{PhysRevLett.121.052003}.

We see that our $M_\Omega^{(3/2)^-}$ data is very consistent with the $\Omega(2012)^-$ and that there are few discernible discretisation effects when converted to physical units; this would be noticeably more severe if the scales from \cite{RQCD:2022xux} or \cite{Bruno:2016plf} were used. Unlike for the $M_\Omega^{(3/2)^+}$ there are obvious discrepancies between X451 and N451, and H105 and N101, suggesting there is some non-negligible pion-based finite volume effect as well as one from the kaon; by eye these point in the same direction (reducing the mass in the infinite volume). Without having a rigorous fit-form at N$^3$LO we leave this result as a prediction of the quantum numbers of the $\Omega(2012)^-$ being $I(J^P)=0(3/2)^-$, as indicated in prior lattice studies \cite{Engel:2013ig}.

%% file: conclusions.tex
\section{Conclusions}

We have shown the viability of very high-precision scale-setting in lattice QCD via the positive-parity ground-state $\Omega$-baryon mass using a combination of gauge-fixed wall sources and the generalised Pencil of Functions technique. This allowed us to fit our data unambiguously at early times allowing for a near-unprecedented level of statistical resolution on the masses of the $\Omega$, below $0.1\%$ relative statistical uncertainty for our largest boxes. This is thanks to the large volume average available from the Coulomb gauge-fixed wall sources and our liberal use of the truncated-solver method.

Finding a reasonable chiral description of such precise data was a challenge, with the upshot that expressions at N$^3$LO in flavor-$\text{SU}(3)$ chiral perturbation theory were needed. Our results for the lattice scales are pretty close in value and precision to previous N$^3$LO results, which were based on a global fit of baryon octet and decuplet masses on mostly the same CLS ensembles but with a different data set. The unexpected success of our chiral flavor-$\text{SU}(3)$ extrapolation of the $\Omega$-mass suggests an extension of the analysis to additional CLS ensembles, in order to obtain more constraints on the LECs of meson-baryon systems.

Such precision scale setting is urgently needed for lattice determinations of quantities of current particular interest such as the leading-order hadronic contribution to the muon g-2, which the community aims to resolve below the percent level including all systematics. Another avenue where such precision is warranted is in the tuning of heavy quark actions for the charm or bottom, where the overall scale ambiguity is large in comparison to the experimentally-determined mass-uncertainty of the states. We have achieved a much more precise determination of the intermediate scale-setting parameter $\sqrt{t_0}$ as well as lattice spacings with relative uncertainties at the level of $0.17\% \dots 0.32 \%$. Our lattice spacing determination on the finest ensembles ($\beta=3.85$) is the least-precise and an ensemble at lighter pion mass, with large physical volume at this $\beta$ would be needed to reduce this uncertainty. Otherwise, larger boxes closer to the physical pion and kaon mass in combination with better resolution of these masses and $t_0$ will be necessary to improve these scale determinations.

We close by illustrating our new lattice scale determinations being applied to our negative-parity $\Omega$-baryon masses. We find that, upon converting to physical units, no discernible cut-off dependence is seen, and that we obtain a state with the quantum numbers $I(J^P)=0(3/2^-)$ consistent with the experimental $\Omega(2012)^-$.

%% file: appendices.tex
\section{Further chiral forms}\label{app:chiral_extra}

Here we detail the various tadpole functions, $I_Q^{(0)}$ , $\bar I_Q^{(2)}$ and bubble functions $J_{K \Xi}(M_\Omega)$, $J_{\eta \Omega}(M_\Omega)$, and $J_{K \Xi^*}(M_\Omega)$ with their specific dependencies on the hadron masses and lattice extents. The tadpole and bubble loop functions are
\allowdisplaybreaks[1]
\begin{eqnarray}
&& \bar I^{(0)}_Q =\frac{m_Q^2}{(4\,\pi)^2}\, \Big\{
\log \left( \frac{m_Q^2}{\mu^2}\right) + \frac{4}{L\,m_Q}\,\sum^{\vec n \neq 0}_{\vec n \in  Z^3} \frac{1}{|\vec n|}\,K_1(|\vec n| \,m_Q\,L) \Big\}\,
\nonumber\\
&& \bar I^{(2)}_Q = \frac{1}{4}\frac{m_Q^4}{(4\,\pi)^2} \Big\{
\log \left( \frac{m_Q^2}{\mu^2}\right) - \frac{16}{(L\,m_Q)^2}\,\sum^{\vec n \neq 0}_{\vec n \in  Z^3} \frac{1}{|\vec n|^2}\,K_2(|\vec n|\,m_Q\,L) \Big\}\,,
\nonumber\\ \nonumber\\
&&J^{}_{K \Xi}(M_\Omega) =
 \frac{(M_\Xi+M_\Omega)^2}{24\,M_{\Omega }^3}\,\Big(M^2_\Xi-M_{\Omega }^2\Big)\,\Big( \bar I^{(0)}_K - m_K^2\,\bar I^{(0)}_\Xi/M_\Xi^2 \Big)
\nonumber\\
&& \qquad \qquad \qquad 
-\,\frac{1}{3}\,\big( E_\Xi +M_\Xi\big)\,p_{K \Xi}^{\,2}\,
\bar I_{K\Xi}(M_\Omega )+ \frac{2}{3}\,\alpha_{K \Xi} \,,
\nonumber\\
&&J_{K \Xi^*}(M_\Omega )=
- \frac{ M_\Omega^2 + M_{\Xi^*}^2}{18\,M_{\Omega }^2\,M_{\Xi^*}}\,\Big(M^2_{\Xi^*}-M_{\Omega }^2\Big)\,\Big( \bar I^{(0)}_K - m_K^2\,\bar I^{(0)}_{\Xi^*}/M_{\Xi^*}^2 \Big)
\nonumber\\
&& \qquad \qquad \qquad 
+\,\frac{M_{\Xi^*}^4+M_\Omega^4  + 12\,M_{\Xi^*}^2\,M_\Omega^2 }{36\,M_{\Omega }^3\,M_{\Xi^*}^2}\,\Big(M^2_{\Xi^*}-M_{\Omega }^2\Big)\,\Big( \bar I^{(0)}_K - m_Q^2\,\bar I^{(0)}_{\Xi^*}/M_{\Xi^*}^2 \Big)
\nonumber\\
&& \qquad \qquad \qquad 
-\,\frac{(M_\Omega+M_{\Xi^*})^2}{9\,M_{\Xi^*}^2}\,\frac{2\,E_{\Xi^*}\,(E_{\Xi^*}-M_{\Xi^*})+5\,M_{\Xi^*}^2}{E_{\Xi^*}+M_{\Xi^*}}\,
p_{K \Xi^*}^{\,2}\,\bar I_{K\Xi^*}(M_\Omega ) \,,
\nonumber\\
&& p_{Q R}^2 =
\frac{M_\Omega^2}{4}-\frac{M_R^2+m_Q^2}{2}+\frac{(M_R^2-m_Q^2)^2}{4\,M_\Omega^2} \,,\qquad \qquad 
E_R^2=M_R^2+p_{QR}^2 \,,
\label{recall-loops}
\end{eqnarray}
with a finite-box scalar bubble loop function $ \bar I_{Q R}(M_\Omega )$ detailed below. We note that our expression (\ref{res-MOmega}) does not depend on the renormalization scale $\mu$, as any dependence is balanced by the counter terms $e_\Omega^{(Q)}$. 
The additional subtraction term $\alpha^{(\Omega)}_{QR}$ in (\ref{recall-loops}) has various implications. It prevents a renormalization of the LEC $d_0$ and $d_D$, but also leads to the wave-function renormalization factor  $Z_\Omega $ of one in the chiral limit. We truncate the sums in $I_Q^{(0/2)}$ to the first twelve lattice displacements $\vec n$.

We recall the in-box form of scalar bubble from \cite{Lutz:2014oxa}
\begin{eqnarray}
&& \bar I_{Q R}(M_\Omega )=  \frac{1}{16\,\pi^2}
\left\{ \gamma^R_{\Omega} + \frac{1}{2}\, \left(\frac{M_R^2-m_Q^2}{M_\Omega^2}-1
\right)
\,\log \left( \frac{m_Q^2}{M_R^2}\right)
\right.
\nonumber\\
&& +\left.
\frac{p_{Q R}}{M_\Omega}\,
\left( \log \left(1-\frac{M_\Omega^2-2\,p_{Q R}\,M_\Omega}{m_Q^2+M_R^2} \right)
-\log \left(1-\frac{M_\Omega^2+2\,p_{Q R}\,M_\Omega}{m_Q^2+M_R^2} \right)\right)
\right\}
\nonumber\\
&&  +\,
\frac{1}{8\,\pi^2}\sum^{\vec n \neq 0}_{\vec n \in  Z^3}
\Big( \int_0^1 d z K_0(|\vec n|\, \mu_{QR} (z) \,L) - \frac{2 \,m_Q\,L \, K_1(|\vec n|\,m_Q \,L)}{|\vec n|\,(M_R^2- m_Q^2) \,L^2} \Big) \;,
\nonumber\\
&& {\rm with } \qquad  \gamma^\Xi_{\Omega} = -  \frac{M^2-(M + \Delta)^2}{(M + \Delta)^2}\,\log \left|\frac{M^2-(M + \Delta)^2}{M^2}\right| \,, \qquad \gamma_\Omega^{\Omega, \Xi^*} =0\,,
\nonumber\\
&& \qquad \qquad \mu_{QR}^2(z) = z\,M_R^2+(1-z)\,m_Q^2- (1-z)\,z\,M_\Omega^2 \;,
\end{eqnarray}
where in our work we have $M_\Omega < m_Q+ M_R$ always. The relative squared momentum $p^2_{QR}(M_\Omega^2)< 0$ was already recalled in (\ref{recall-loops}).
The subtraction term in (\ref{recall-loops}) is specified with
\begin{eqnarray}
&& \alpha^{(\Omega)}_{K \Xi} = \frac{\beta_1\,\Delta^2}{(4\,\pi)^2} \Bigg\{ 
\Big( M_\Omega - M - \Delta\Big)\, \Big( \frac{\Delta\,\partial}{\partial\,\Delta} + 1\Big)
\nonumber\\
&& \qquad  -\,\Big( M_\Xi - M \Big)\, \Big( \frac{\Delta\,\partial}{\partial\,\Delta} - \frac{\Delta\,\partial}{\partial\,M} 
+ \frac{M+ \Delta}{M}\Big)
\Bigg\}\,\delta_1 
 + \frac{\Delta\,m_K^2}{(4\,\pi)^2}\,\beta_1\,\delta_2 \,, 
\nonumber\\ \nonumber\\
&&\beta_1 = \frac{(2\,M+\Delta)^4}{16\,M\,(M+\Delta)^3}\,, \qquad \qquad \delta_1 = -\frac{M\,(2\,M+ \Delta)}{(M+\Delta)^2}\,\log \frac{\Delta\,(2\,M+\Delta)}{M^2}\,, \qquad  
\nonumber\\
&& \delta_2 = \frac{M}{2\,M+ \Delta } +M\,\frac{2\,M^2 + 2\,\Delta\,M+\Delta^2}{(2\,M + \Delta )\,(M+ \Delta )^2}\,
\log \frac{\Delta\,(2\,M+ \Delta)}{M^2}  \,,
\label{def-alphaBR}
\end{eqnarray}
in terms of dimensionless parameters $\beta_n$ and $\delta_n$ depending  on the ratio $\Delta/M$ only. They are derived in Appendix A and Appendix B of \cite{Lutz:2018cqo}. 
Here  $M+ \Delta$ gives the chiral limit value of the $\Omega$ and $\Xi^*$, the value of $M$ that limit of the $\Xi$ mass.
While the $\alpha_n $ and $\beta_n $ are rational functions in $\Delta/M$ 
properly normalized to one in the limit $\Delta \to 0$, the  $\delta_n $
have a more complicated form involving terms proportional to $\log (\Delta/M)$.  As was emphasized in \cite{Lutz:2018cqo} an expansion of such coefficients in powers of $\Delta/M$ is futile if truncated at low orders. At realistic values for $\Delta$ and $M$ the coefficients $\beta_n$ depart strongly from their limit value one, with even flipped signs for some cases. 

A matching to the LECs of the chiral Lagrangians leads to  
\begin{eqnarray}
&& g^{(S)}_{\Omega \pi} = 3\,\hat h^{(S)}_1 /4 -  7\,d_0 + d_D \,,\!
\qquad \qquad \qquad \quad 
g^{(V)}_{ \Omega \pi} = \frac{3}{4}\,( M + \Delta)\,\hat h^{(V)}_1 \,,
\nonumber\\
&& g^{(S)}_{\Omega K} =  \hat h^{(S)}_1 +\hat h^{(S)}_2+\hat h^{(S)}_3 - 8\,d_0 -4\, d_D  \,,\qquad \;\;\!\!\!
g^{(V)}_{ \Omega K} =(M + \Delta)\, \big(\hat h^{(V)}_1 +\hat h^{(V)}_2+\hat h^{V}_3 \big) \,,
\nonumber\\
&& g^{(S)}_{ \Omega \eta } = \hat h^{(S)}_1/4 +\hat h^{(S)}_2/3 - 3\,  d_0 -3\,d_D \,,\!\!\! \!\! 
\qquad \quad \;\;\;
g^{(V)}_{\Omega \eta } =(M + \Delta)\, (\hat h^{(V)}_1/4 +\hat h^{(V)}_2/3)\,,
\nonumber\\
&& \hat h_1^{(S)}= h_1^{(S)} + h_2^{(S)}/3 \,,\qquad \qquad 
 \hat h_1^{(V)}= h_1^{(V)} - \frac{1}{3}\,h_2^{(S)}/(M+\Delta ) \,,
\nonumber\\
&& \hat h_2^{(S)}= h_3^{(S)} + h_4^{(S)}/3 \,, \qquad \qquad 
\hat h_2^{(V)}= h_2^{(V)} -  \frac{1}{3}\,h_4^{(S)}/(M+\Delta )\,,
\nonumber\\
&& \hat h_3^{(S)}= h_5^{(S)} + h_6^{(S)}/3\,, \qquad \qquad 
\hat h_3^{(V)}= h_3^{(V)} -  \frac{1}{3}\,h_6^{(S)}/(M+\Delta )\,,
\nonumber\\
&& e^{(\pi )} _\Omega =\frac{1}{5}\,\big(-11 \,e_0 - e_2 + 5 \,e_3 - 9 \,e_4\big)
+16\,\frac{d_0-d_D}{f^2}\,(2\,L_8 -L_5)
\nonumber\\
&& \qquad \!+\,
\frac{16}{5}\,\frac{9\,d_0-5\,d_D}{f^2}\,(2\,L_6 -L_4)\,,  
\nonumber\\
&& e^{(K)} _\Omega =- \frac{4}{5}\,\Big(
e_0 + e_2 + 5\, e_3 + 9\, e_4 \Big)
+32\,\frac{d_0+d_D}{f^2}\,(2\,L_8 -L_5)+ 
\frac{64}{5}\,\frac{9\,d_0+5\,d_D}{f^2}\,(2\,L_6 -L_4)\,,
\nonumber\\
&& e^{(\eta )} _\Omega = -\frac{12}{5}\,\Big( e_0 + e_2 - e_4\Big) +\frac{192}{5}\,\frac{d_0}{f^2}\,(2\,L_6 -L_4)\,.
\end{eqnarray}

\section{Lattice fit results\label{app:lattice_fits}}

\begin{table}[h!]
\begin{tabular}{@{\hspace{1em}}c@{\hspace{1em}}|@{\hspace{1em}}c@{\hspace{1em}}|@{\hspace{1em}}c@{\hspace{1em}}|@{\hspace{1em}}c@{\hspace{1em}}}
\toprule
$a(\beta=3.34)$ & $0.4728(14)(02) \text{ GeV}^{-1}$ & $a(\beta=3.40)$ & $0.4187(09)(02) \text{ GeV}^{-1}$ \\
$a(\beta=3.46)$ & $0.3740(07)(02) \text{ GeV}^{-1}$ & $a(\beta=3.55)$ & $0.3176(05)(02) \text{ GeV}^{-1}$ \\
$a(\beta=3.70)$ & $0.2475(06)(01) \text{ GeV}^{-1}$ & $a(\beta=3.85)$ & $0.1929(06)(01) \text{ GeV}^{-1}$ \\
\hline
$\tilde{d}_0$ & $-0.054(18)$ & $\tilde{d}_D$ & $-0.071(20)$ \\
$\tilde{c}_\Omega$ & $0.91(22)$ & $\tilde{h}_\Omega$ & $0.34(8)$ \\
$\tilde{g}_{\Omega K}^{(S)}$ & $-1.8(5)$ & $\tilde{g}_{\Omega K}^{(V)}$ & $6.6(1.7)$ \\
$\tilde{e}_\Omega^{(\eta)}$ & $-0.035(10)$ & $t_0$ & $0.5385(24)(5) \text{ GeV}^{-2}$ \\
\botrule
\end{tabular}
\caption{Our final lattice fit results in the tilded lattice parameters. All other parameters in Eq.~\ref{res-MOmega-Lattice} are set to zero.}\label{tab:finalfit}
\end{table}

Here we give the initial fit results in terms of the lattice (tilded) parameters. These are tabulated in Tab.~\ref{tab:finalfit}.

\section{Correlation matrix}

\begin{scriptsize}
\begin{table}[h!]
\begin{tabular}{c|ccccccc}
&
$a(\beta=3.34)$ & 
$a(\beta=3.40)$ & 
$a(\beta=3.46)$ & 
$a(\beta=3.55)$ & 
$a(\beta=3.70)$ & 
$a(\beta=3.85)$ &  
$t_0$ \\
\hline

$a(\beta=3.34)$ & 1 & 0.8300 & 0.8505 & 0.8211 & 0.6577 & 0.6214 & 0.7022 \\
$a(\beta=3.40)$ & & 1 & 0.9360 & 0.9222 & 0.6424 & 0.7368 & 0.8725 \\
$a(\beta=3.46)$ & & & 1 &  0.9546 & 0.7107 & 0.7389 & 0.9081 \\
$a(\beta=3.55)$ & & & & 1 & 0.7238 & 0.7250 & 0.9398 \\
$a(\beta=3.70)$ & & & & & 1 &  0.4980 & 0.7166 \\
$a(\beta=3.85)$ &  &  &  &  &  & 1 & 0.7254 \\
$t_0$ & & & & & & & 1 \\
\end{tabular}
\caption{Normalised correlation matrix for our lattice spacings and $t_0$.}\label{tab:latcorr}
\end{table}
\end{scriptsize}

The correlation matrix between the values for the lattice spacings and the intermediate scale $t_0$ for our final results in physical units from Tab.~\ref{tab:finalfit_fermi}
is listed in Tab.~\ref{tab:latcorr}.

\section{Extra measurements}\label{app:further_measurements}

\begin{table}[h!]
\begin{tabular}{c|cccccc}
\toprule
id & $L^3\times T$ & $am_\pi$ & $am_K$ & $am_\Xi$ & $am_{\Xi^*}$ & $t_0/a^2$ \\
\hline
B650 & $32^3\times 64$ & 0.1283(12) & 0.2319(6) & 0.6109(17) & 0.727(5) & 2.232(6) \\
\hline
X451 & $40^3\times 128$ & 0.11047(38) & 0.17803(17) & 0.4799(11) & 0.572(5) & 3.688(4) \\
N451 & $48^3\times 128$ & 0.11072(29) \cite{Ce:2022kxy} & 0.17824(18) \cite{Ce:2022kxy} & 0.4799(11) & 0.572(5) & 3.681(7) \cite{Ce:2022kxy} \\
D452 & $64^3\times 128$ & 0.05941(55) \cite{Ce:2022kxy} & 0.18651(15) \cite{Ce:2022kxy} & 0.4942(8) & 0.573(5) & 3.725(1) \cite{Ce:2022kxy} \\
\botrule
\end{tabular}
\caption{A list of quantities needed for our fit that (partially, in some cases) don't yet appear in the literature.}\label{tab:extra_meas}
\end{table}

In this work we have used several ensembles that have not had basic measurements of the spectrum performed on them. Here we fill in those gaps and provide determinations of $\pi$, K, $\Xi$, and $\Xi^*$ masses. Here, the values for $\Xi^*$ only come from estimates based on similar ensembles and extrapolations of the \cite{RQCD:2022xux} data.

%% file: omega.bbl
\begin{thebibliography}{54}%
\makeatletter
\providecommand \@ifxundefined [1]{%
 \@ifx{#1\undefined}
}%
\providecommand \@ifnum [1]{%
 \ifnum #1\expandafter \@firstoftwo
 \else \expandafter \@secondoftwo
 \fi
}%
\providecommand \@ifx [1]{%
 \ifx #1\expandafter \@firstoftwo
 \else \expandafter \@secondoftwo
 \fi
}%
\providecommand \natexlab [1]{#1}%
\providecommand \enquote  [1]{``#1''}%
\providecommand \bibnamefont  [1]{#1}%
\providecommand \bibfnamefont [1]{#1}%
\providecommand \citenamefont [1]{#1}%
\providecommand \href@noop [0]{\@secondoftwo}%
\providecommand \href [0]{\begingroup \@sanitize@url \@href}%
\providecommand \@href[1]{\@@startlink{#1}\@@href}%
\providecommand \@@href[1]{\endgroup#1\@@endlink}%
\providecommand \@sanitize@url [0]{\catcode `\\12\catcode `\$12\catcode `\&12\catcode `\#12\catcode `\^12\catcode `\_12\catcode `\%12\relax}%
\providecommand \@@startlink[1]{}%
\providecommand \@@endlink[0]{}%
\providecommand \url  [0]{\begingroup\@sanitize@url \@url }%
\providecommand \@url [1]{\endgroup\@href {#1}{\urlprefix }}%
\providecommand \urlprefix  [0]{URL }%
\providecommand \Eprint [0]{\href }%
\providecommand \doibase [0]{http://dx.doi.org/}%
\providecommand \selectlanguage [0]{\@gobble}%
\providecommand \bibinfo  [0]{\@secondoftwo}%
\providecommand \bibfield  [0]{\@secondoftwo}%
\providecommand \translation [1]{[#1]}%
\providecommand \BibitemOpen [0]{}%
\providecommand \bibitemStop [0]{}%
\providecommand \bibitemNoStop [0]{.\EOS\space}%
\providecommand \EOS [0]{\spacefactor3000\relax}%
\providecommand \BibitemShut  [1]{\csname bibitem#1\endcsname}%
\let\auto@bib@innerbib\@empty
\bibitem [{\citenamefont {Aoyama}\ \emph {et~al.}(2020)\citenamefont {Aoyama} \emph {et~al.}}]{Aoyama:2020ynm}%
  \BibitemOpen
  \bibfield  {author} {\bibinfo {author} {\bibfnamefont {T.}~\bibnamefont {Aoyama}} \emph {et~al.},\ }\href {\doibase 10.1016/j.physrep.2020.07.006} {\bibfield  {journal} {\bibinfo  {journal} {Phys. Rept.}\ }\textbf {\bibinfo {volume} {887}},\ \bibinfo {pages} {1} (\bibinfo {year} {2020})},\ \Eprint {http://arxiv.org/abs/2006.04822} {arXiv:2006.04822 [hep-ph]} \BibitemShut {NoStop}%
\bibitem [{\citenamefont {Borsanyi}\ \emph {et~al.}(2021)\citenamefont {Borsanyi} \emph {et~al.}}]{Borsanyi:2020mff}%
  \BibitemOpen
  \bibfield  {author} {\bibinfo {author} {\bibfnamefont {S.}~\bibnamefont {Borsanyi}} \emph {et~al.},\ }\href {\doibase 10.1038/s41586-021-03418-1} {\bibfield  {journal} {\bibinfo  {journal} {Nature}\ }\textbf {\bibinfo {volume} {593}},\ \bibinfo {pages} {51} (\bibinfo {year} {2021})},\ \Eprint {http://arxiv.org/abs/2002.12347} {arXiv:2002.12347 [hep-lat]} \BibitemShut {NoStop}%
\bibitem [{\citenamefont {Della~Morte}\ \emph {et~al.}(2017)\citenamefont {Della~Morte}, \citenamefont {Francis}, \citenamefont {G\"ulpers}, \citenamefont {Herdo\'\i{}za}, \citenamefont {von Hippel}, \citenamefont {Horch}, \citenamefont {J\"ager}, \citenamefont {Meyer}, \citenamefont {Nyffeler},\ and\ \citenamefont {Wittig}}]{DellaMorte:2017dyu}%
  \BibitemOpen
  \bibfield  {author} {\bibinfo {author} {\bibfnamefont {M.}~\bibnamefont {Della~Morte}}, \bibinfo {author} {\bibfnamefont {A.}~\bibnamefont {Francis}}, \bibinfo {author} {\bibfnamefont {V.}~\bibnamefont {G\"ulpers}}, \bibinfo {author} {\bibfnamefont {G.}~\bibnamefont {Herdo\'\i{}za}}, \bibinfo {author} {\bibfnamefont {G.}~\bibnamefont {von Hippel}}, \bibinfo {author} {\bibfnamefont {H.}~\bibnamefont {Horch}}, \bibinfo {author} {\bibfnamefont {B.}~\bibnamefont {J\"ager}}, \bibinfo {author} {\bibfnamefont {H.~B.}\ \bibnamefont {Meyer}}, \bibinfo {author} {\bibfnamefont {A.}~\bibnamefont {Nyffeler}}, \ and\ \bibinfo {author} {\bibfnamefont {H.}~\bibnamefont {Wittig}},\ }\href {\doibase 10.1007/JHEP10(2017)020} {\bibfield  {journal} {\bibinfo  {journal} {JHEP}\ }\textbf {\bibinfo {volume} {10}},\ \bibinfo {pages} {020} (\bibinfo {year} {2017})},\ \Eprint {http://arxiv.org/abs/1705.01775} {arXiv:1705.01775 [hep-lat]} \BibitemShut {NoStop}%
\bibitem [{\citenamefont {Bazavov}\ \emph {et~al.}(2012)\citenamefont {Bazavov} \emph {et~al.}}]{FermilabLattice:2011njy}%
  \BibitemOpen
  \bibfield  {author} {\bibinfo {author} {\bibfnamefont {A.}~\bibnamefont {Bazavov}} \emph {et~al.} (\bibinfo {collaboration} {Fermilab Lattice, MILC}),\ }\href {\doibase 10.1103/PhysRevD.85.114506} {\bibfield  {journal} {\bibinfo  {journal} {Phys. Rev. D}\ }\textbf {\bibinfo {volume} {85}},\ \bibinfo {pages} {114506} (\bibinfo {year} {2012})},\ \Eprint {http://arxiv.org/abs/1112.3051} {arXiv:1112.3051 [hep-lat]} \BibitemShut {NoStop}%
\bibitem [{\citenamefont {Bruno}\ \emph {et~al.}(2017)\citenamefont {Bruno}, \citenamefont {Korzec},\ and\ \citenamefont {Schaefer}}]{Bruno:2016plf}%
  \BibitemOpen
  \bibfield  {author} {\bibinfo {author} {\bibfnamefont {M.}~\bibnamefont {Bruno}}, \bibinfo {author} {\bibfnamefont {T.}~\bibnamefont {Korzec}}, \ and\ \bibinfo {author} {\bibfnamefont {S.}~\bibnamefont {Schaefer}},\ }\href {\doibase 10.1103/PhysRevD.95.074504} {\bibfield  {journal} {\bibinfo  {journal} {Phys. Rev. D}\ }\textbf {\bibinfo {volume} {95}},\ \bibinfo {pages} {074504} (\bibinfo {year} {2017})},\ \Eprint {http://arxiv.org/abs/1608.08900} {arXiv:1608.08900 [hep-lat]} \BibitemShut {NoStop}%
\bibitem [{\citenamefont {Bali}\ \emph {et~al.}(2013)\citenamefont {Bali} \emph {et~al.}}]{Bali:2012qs}%
  \BibitemOpen
  \bibfield  {author} {\bibinfo {author} {\bibfnamefont {G.~S.}\ \bibnamefont {Bali}} \emph {et~al.},\ }\href {\doibase 10.1016/j.nuclphysb.2012.08.009} {\bibfield  {journal} {\bibinfo  {journal} {Nucl. Phys. B}\ }\textbf {\bibinfo {volume} {866}},\ \bibinfo {pages} {1} (\bibinfo {year} {2013})},\ \Eprint {http://arxiv.org/abs/1206.7034} {arXiv:1206.7034 [hep-lat]} \BibitemShut {NoStop}%
\bibitem [{\citenamefont {Bali}\ \emph {et~al.}(2023)\citenamefont {Bali}, \citenamefont {Collins}, \citenamefont {Georg}, \citenamefont {Jenkins}, \citenamefont {Korcyl}, \citenamefont {Sch\"afer}, \citenamefont {Scholz}, \citenamefont {Simeth}, \citenamefont {S\"oldner},\ and\ \citenamefont {Weish\"aupl}}]{RQCD:2022xux}%
  \BibitemOpen
  \bibfield  {author} {\bibinfo {author} {\bibfnamefont {G.~S.}\ \bibnamefont {Bali}}, \bibinfo {author} {\bibfnamefont {S.}~\bibnamefont {Collins}}, \bibinfo {author} {\bibfnamefont {P.}~\bibnamefont {Georg}}, \bibinfo {author} {\bibfnamefont {D.}~\bibnamefont {Jenkins}}, \bibinfo {author} {\bibfnamefont {P.}~\bibnamefont {Korcyl}}, \bibinfo {author} {\bibfnamefont {A.}~\bibnamefont {Sch\"afer}}, \bibinfo {author} {\bibfnamefont {E.~E.}\ \bibnamefont {Scholz}}, \bibinfo {author} {\bibfnamefont {J.}~\bibnamefont {Simeth}}, \bibinfo {author} {\bibfnamefont {W.}~\bibnamefont {S\"oldner}}, \ and\ \bibinfo {author} {\bibfnamefont {S.}~\bibnamefont {Weish\"aupl}} (\bibinfo {collaboration} {RQCD}),\ }\href {\doibase 10.1007/JHEP05(2023)035} {\bibfield  {journal} {\bibinfo  {journal} {JHEP}\ }\textbf {\bibinfo {volume} {05}},\ \bibinfo {pages} {035} (\bibinfo {year} {2023})},\ \Eprint {http://arxiv.org/abs/2211.03744} {arXiv:2211.03744 [hep-lat]} \BibitemShut {NoStop}%
\bibitem [{\citenamefont {Blum}\ \emph {et~al.}(2016)\citenamefont {Blum} \emph {et~al.}}]{RBC:2014ntl}%
  \BibitemOpen
  \bibfield  {author} {\bibinfo {author} {\bibfnamefont {T.}~\bibnamefont {Blum}} \emph {et~al.} (\bibinfo {collaboration} {RBC, UKQCD}),\ }\href {\doibase 10.1103/PhysRevD.93.074505} {\bibfield  {journal} {\bibinfo  {journal} {Phys. Rev. D}\ }\textbf {\bibinfo {volume} {93}},\ \bibinfo {pages} {074505} (\bibinfo {year} {2016})},\ \Eprint {http://arxiv.org/abs/1411.7017} {arXiv:1411.7017 [hep-lat]} \BibitemShut {NoStop}%
\bibitem [{\citenamefont {Miller}\ \emph {et~al.}(2021)\citenamefont {Miller} \emph {et~al.}}]{Miller:2020evg}%
  \BibitemOpen
  \bibfield  {author} {\bibinfo {author} {\bibfnamefont {N.}~\bibnamefont {Miller}} \emph {et~al.},\ }\href {\doibase 10.1103/PhysRevD.103.054511} {\bibfield  {journal} {\bibinfo  {journal} {Phys. Rev. D}\ }\textbf {\bibinfo {volume} {103}},\ \bibinfo {pages} {054511} (\bibinfo {year} {2021})},\ \Eprint {http://arxiv.org/abs/2011.12166} {arXiv:2011.12166 [hep-lat]} \BibitemShut {NoStop}%
\bibitem [{\citenamefont {Davies}\ \emph {et~al.}(2004)\citenamefont {Davies} \emph {et~al.}}]{HPQCD:2003rsu}%
  \BibitemOpen
  \bibfield  {author} {\bibinfo {author} {\bibfnamefont {C.~T.~H.}\ \bibnamefont {Davies}} \emph {et~al.} (\bibinfo {collaboration} {HPQCD, UKQCD, MILC, Fermilab Lattice}),\ }\href {\doibase 10.1103/PhysRevLett.92.022001} {\bibfield  {journal} {\bibinfo  {journal} {Phys. Rev. Lett.}\ }\textbf {\bibinfo {volume} {92}},\ \bibinfo {pages} {022001} (\bibinfo {year} {2004})},\ \Eprint {http://arxiv.org/abs/hep-lat/0304004} {arXiv:hep-lat/0304004} \BibitemShut {NoStop}%
\bibitem [{\citenamefont {Necco}\ and\ \citenamefont {Sommer}(2002)}]{Necco:2001xg}%
  \BibitemOpen
  \bibfield  {author} {\bibinfo {author} {\bibfnamefont {S.}~\bibnamefont {Necco}}\ and\ \bibinfo {author} {\bibfnamefont {R.}~\bibnamefont {Sommer}},\ }\href {\doibase 10.1016/S0550-3213(01)00582-X} {\bibfield  {journal} {\bibinfo  {journal} {Nucl. Phys. B}\ }\textbf {\bibinfo {volume} {622}},\ \bibinfo {pages} {328} (\bibinfo {year} {2002})},\ \Eprint {http://arxiv.org/abs/hep-lat/0108008} {arXiv:hep-lat/0108008} \BibitemShut {NoStop}%
\bibitem [{\citenamefont {Bernard}\ \emph {et~al.}(2000)\citenamefont {Bernard}, \citenamefont {Burch}, \citenamefont {Orginos}, \citenamefont {Toussaint}, \citenamefont {DeGrand}, \citenamefont {DeTar}, \citenamefont {Gottlieb}, \citenamefont {Heller}, \citenamefont {Hetrick},\ and\ \citenamefont {Sugar}}]{Bernard:2000gd}%
  \BibitemOpen
  \bibfield  {author} {\bibinfo {author} {\bibfnamefont {C.~W.}\ \bibnamefont {Bernard}}, \bibinfo {author} {\bibfnamefont {T.}~\bibnamefont {Burch}}, \bibinfo {author} {\bibfnamefont {K.}~\bibnamefont {Orginos}}, \bibinfo {author} {\bibfnamefont {D.}~\bibnamefont {Toussaint}}, \bibinfo {author} {\bibfnamefont {T.~A.}\ \bibnamefont {DeGrand}}, \bibinfo {author} {\bibfnamefont {C.~E.}\ \bibnamefont {DeTar}}, \bibinfo {author} {\bibfnamefont {S.~A.}\ \bibnamefont {Gottlieb}}, \bibinfo {author} {\bibfnamefont {U.~M.}\ \bibnamefont {Heller}}, \bibinfo {author} {\bibfnamefont {J.~E.}\ \bibnamefont {Hetrick}}, \ and\ \bibinfo {author} {\bibfnamefont {B.}~\bibnamefont {Sugar}},\ }\href {\doibase 10.1103/PhysRevD.62.034503} {\bibfield  {journal} {\bibinfo  {journal} {Phys. Rev. D}\ }\textbf {\bibinfo {volume} {62}},\ \bibinfo {pages} {034503} (\bibinfo {year} {2000})},\ \Eprint {http://arxiv.org/abs/hep-lat/0002028} {arXiv:hep-lat/0002028} \BibitemShut {NoStop}%
\bibitem [{\citenamefont {L\"uscher}(2010)}]{Luscher:2010iy}%
  \BibitemOpen
  \bibfield  {author} {\bibinfo {author} {\bibfnamefont {M.}~\bibnamefont {L\"uscher}},\ }\href {\doibase 10.1007/JHEP08(2010)071} {\bibfield  {journal} {\bibinfo  {journal} {JHEP}\ }\textbf {\bibinfo {volume} {08}},\ \bibinfo {pages} {071} (\bibinfo {year} {2010})},\ \bibinfo {note} {[Erratum: JHEP 03, 092 (2014)]},\ \Eprint {http://arxiv.org/abs/1006.4518} {arXiv:1006.4518 [hep-lat]} \BibitemShut {NoStop}%
\bibitem [{\citenamefont {Bors\'anyi}\ \emph {et~al.}(2012)\citenamefont {Bors\'anyi}, \citenamefont {D\"urr}, \citenamefont {Fodor}, \citenamefont {Hoelbling}, \citenamefont {Katz}, \citenamefont {Krieg}, \citenamefont {Kurth}, \citenamefont {Lellouch}, \citenamefont {Lippert},\ and\ \citenamefont {McNeile}}]{BMW:2012hcm}%
  \BibitemOpen
  \bibfield  {author} {\bibinfo {author} {\bibfnamefont {S.}~\bibnamefont {Bors\'anyi}}, \bibinfo {author} {\bibfnamefont {S.}~\bibnamefont {D\"urr}}, \bibinfo {author} {\bibfnamefont {Z.}~\bibnamefont {Fodor}}, \bibinfo {author} {\bibfnamefont {C.}~\bibnamefont {Hoelbling}}, \bibinfo {author} {\bibfnamefont {S.~D.}\ \bibnamefont {Katz}}, \bibinfo {author} {\bibfnamefont {S.}~\bibnamefont {Krieg}}, \bibinfo {author} {\bibfnamefont {T.}~\bibnamefont {Kurth}}, \bibinfo {author} {\bibfnamefont {L.}~\bibnamefont {Lellouch}}, \bibinfo {author} {\bibfnamefont {T.}~\bibnamefont {Lippert}}, \ and\ \bibinfo {author} {\bibfnamefont {C.}~\bibnamefont {McNeile}} (\bibinfo {collaboration} {BMW}),\ }\href {\doibase 10.1007/JHEP09(2012)010} {\bibfield  {journal} {\bibinfo  {journal} {JHEP}\ }\textbf {\bibinfo {volume} {09}},\ \bibinfo {pages} {010} (\bibinfo {year} {2012})},\ \Eprint {http://arxiv.org/abs/1203.4469} {arXiv:1203.4469 [hep-lat]} \BibitemShut {NoStop}%
\bibitem [{\citenamefont {Lutz}\ \emph {et~al.}(2023)\citenamefont {Lutz}, \citenamefont {Heo},\ and\ \citenamefont {Guo}}]{Lutz:2023xpi}%
  \BibitemOpen
  \bibfield  {author} {\bibinfo {author} {\bibfnamefont {M.~F.~M.}\ \bibnamefont {Lutz}}, \bibinfo {author} {\bibfnamefont {Y.}~\bibnamefont {Heo}}, \ and\ \bibinfo {author} {\bibfnamefont {X.-Y.}\ \bibnamefont {Guo}},\ }\href {\doibase 10.1140/epjc/s10052-023-11556-1} {\bibfield  {journal} {\bibinfo  {journal} {Eur. Phys. J. C}\ }\textbf {\bibinfo {volume} {83}},\ \bibinfo {pages} {440} (\bibinfo {year} {2023})},\ \Eprint {http://arxiv.org/abs/2301.06837} {arXiv:2301.06837 [hep-lat]} \BibitemShut {NoStop}%
\bibitem [{\citenamefont {Horsley}\ \emph {et~al.}(2019)\citenamefont {Horsley} \emph {et~al.}}]{CSSM:2019jmq}%
  \BibitemOpen
  \bibfield  {author} {\bibinfo {author} {\bibfnamefont {R.}~\bibnamefont {Horsley}} \emph {et~al.} (\bibinfo {collaboration} {CSSM, QCDSF, UKQCD}),\ }\href {\doibase 10.1088/1361-6471/ab32c1} {\bibfield  {journal} {\bibinfo  {journal} {J. Phys. G}\ }\textbf {\bibinfo {volume} {46}},\ \bibinfo {pages} {115004} (\bibinfo {year} {2019})},\ \Eprint {http://arxiv.org/abs/1904.02304} {arXiv:1904.02304 [hep-lat]} \BibitemShut {NoStop}%
\bibitem [{\citenamefont {Bruno}\ \emph {et~al.}(2015)\citenamefont {Bruno} \emph {et~al.}}]{Bruno:2014jqa}%
  \BibitemOpen
  \bibfield  {author} {\bibinfo {author} {\bibfnamefont {M.}~\bibnamefont {Bruno}} \emph {et~al.},\ }\href {\doibase 10.1007/JHEP02(2015)043} {\bibfield  {journal} {\bibinfo  {journal} {JHEP}\ }\textbf {\bibinfo {volume} {02}},\ \bibinfo {pages} {043} (\bibinfo {year} {2015})},\ \Eprint {http://arxiv.org/abs/1411.3982} {arXiv:1411.3982 [hep-lat]} \BibitemShut {NoStop}%
\bibitem [{\citenamefont {Bali}\ \emph {et~al.}(2016)\citenamefont {Bali}, \citenamefont {Scholz}, \citenamefont {Simeth},\ and\ \citenamefont {S\"oldner}}]{Bali:2016umi}%
  \BibitemOpen
  \bibfield  {author} {\bibinfo {author} {\bibfnamefont {G.~S.}\ \bibnamefont {Bali}}, \bibinfo {author} {\bibfnamefont {E.~E.}\ \bibnamefont {Scholz}}, \bibinfo {author} {\bibfnamefont {J.}~\bibnamefont {Simeth}}, \ and\ \bibinfo {author} {\bibfnamefont {W.}~\bibnamefont {S\"oldner}} (\bibinfo {collaboration} {RQCD}),\ }\href {\doibase 10.1103/PhysRevD.94.074501} {\bibfield  {journal} {\bibinfo  {journal} {Phys. Rev. D}\ }\textbf {\bibinfo {volume} {94}},\ \bibinfo {pages} {074501} (\bibinfo {year} {2016})},\ \Eprint {http://arxiv.org/abs/1606.09039} {arXiv:1606.09039 [hep-lat]} \BibitemShut {NoStop}%
\bibitem [{\citenamefont {Yelton}\ \emph {et~al.}(2018)\citenamefont {Yelton} \emph {et~al.}}]{PhysRevLett.121.052003}%
  \BibitemOpen
  \bibfield  {author} {\bibinfo {author} {\bibfnamefont {J.}~\bibnamefont {Yelton}} \emph {et~al.} (\bibinfo {collaboration} {Belle Collaboration}),\ }\href {\doibase 10.1103/PhysRevLett.121.052003} {\bibfield  {journal} {\bibinfo  {journal} {Phys. Rev. Lett.}\ }\textbf {\bibinfo {volume} {121}},\ \bibinfo {pages} {052003} (\bibinfo {year} {2018})}\BibitemShut {NoStop}%
\bibitem [{Note1()}]{Note1}%
  \BibitemOpen
  \bibinfo {note} {With particular care being taken when going through the temporal boundaries as anti-periodic boundary conditions are applied on the fermions in the solve.}\BibitemShut {Stop}%
\bibitem [{\citenamefont {L{\"u}scher}(2007)}]{Luscher:2007se}%
  \BibitemOpen
  \bibfield  {author} {\bibinfo {author} {\bibfnamefont {M.}~\bibnamefont {L{\"u}scher}},\ }\href {\doibase 10.1088/1126-6708/2007/07/081} {\bibfield  {journal} {\bibinfo  {journal} {JHEP}\ }\textbf {\bibinfo {volume} {07}},\ \bibinfo {pages} {081} (\bibinfo {year} {2007})},\ \Eprint {http://arxiv.org/abs/0706.2298} {arXiv:0706.2298 [hep-lat]} \BibitemShut {NoStop}%
\bibitem [{\citenamefont {Bali}\ \emph {et~al.}(2010)\citenamefont {Bali}, \citenamefont {Collins},\ and\ \citenamefont {Schafer}}]{Bali:2009hu}%
  \BibitemOpen
  \bibfield  {author} {\bibinfo {author} {\bibfnamefont {G.~S.}\ \bibnamefont {Bali}}, \bibinfo {author} {\bibfnamefont {S.}~\bibnamefont {Collins}}, \ and\ \bibinfo {author} {\bibfnamefont {A.}~\bibnamefont {Schafer}},\ }\href {\doibase 10.1016/j.cpc.2010.05.008} {\bibfield  {journal} {\bibinfo  {journal} {Comput. Phys. Commun.}\ }\textbf {\bibinfo {volume} {181}},\ \bibinfo {pages} {1570} (\bibinfo {year} {2010})},\ \Eprint {http://arxiv.org/abs/0910.3970} {arXiv:0910.3970 [hep-lat]} \BibitemShut {NoStop}%
\bibitem [{\citenamefont {Foster}\ and\ \citenamefont {Michael}(1999)}]{Foster:1998vw}%
  \BibitemOpen
  \bibfield  {author} {\bibinfo {author} {\bibfnamefont {M.}~\bibnamefont {Foster}}\ and\ \bibinfo {author} {\bibfnamefont {C.}~\bibnamefont {Michael}} (\bibinfo {collaboration} {UKQCD}),\ }\href {\doibase 10.1103/PhysRevD.59.074503} {\bibfield  {journal} {\bibinfo  {journal} {Phys. Rev. D}\ }\textbf {\bibinfo {volume} {59}},\ \bibinfo {pages} {074503} (\bibinfo {year} {1999})},\ \Eprint {http://arxiv.org/abs/hep-lat/9810021} {arXiv:hep-lat/9810021} \BibitemShut {NoStop}%
\bibitem [{\citenamefont {Boucaud}\ \emph {et~al.}(2008)\citenamefont {Boucaud} \emph {et~al.}}]{ETM:2008zte}%
  \BibitemOpen
  \bibfield  {author} {\bibinfo {author} {\bibfnamefont {P.}~\bibnamefont {Boucaud}} \emph {et~al.} (\bibinfo {collaboration} {ETM}),\ }\href {\doibase 10.1016/j.cpc.2008.06.013} {\bibfield  {journal} {\bibinfo  {journal} {Comput. Phys. Commun.}\ }\textbf {\bibinfo {volume} {179}},\ \bibinfo {pages} {695} (\bibinfo {year} {2008})},\ \Eprint {http://arxiv.org/abs/0803.0224} {arXiv:0803.0224 [hep-lat]} \BibitemShut {NoStop}%
\bibitem [{\citenamefont {Billoire}\ \emph {et~al.}(1985)\citenamefont {Billoire}, \citenamefont {Marinari},\ and\ \citenamefont {Parisi}}]{Billoire:1985yn}%
  \BibitemOpen
  \bibfield  {author} {\bibinfo {author} {\bibfnamefont {A.}~\bibnamefont {Billoire}}, \bibinfo {author} {\bibfnamefont {E.}~\bibnamefont {Marinari}}, \ and\ \bibinfo {author} {\bibfnamefont {G.}~\bibnamefont {Parisi}},\ }\href {\doibase 10.1016/0370-2693(85)91079-2} {\bibfield  {journal} {\bibinfo  {journal} {Phys. Lett. B}\ }\textbf {\bibinfo {volume} {162}},\ \bibinfo {pages} {160} (\bibinfo {year} {1985})}\BibitemShut {NoStop}%
\bibitem [{\citenamefont {Gupta}\ \emph {et~al.}(1991)\citenamefont {Gupta}, \citenamefont {Guralnik}, \citenamefont {Kilcup},\ and\ \citenamefont {Sharpe}}]{Gupta:1990mr}%
  \BibitemOpen
  \bibfield  {author} {\bibinfo {author} {\bibfnamefont {R.}~\bibnamefont {Gupta}}, \bibinfo {author} {\bibfnamefont {G.}~\bibnamefont {Guralnik}}, \bibinfo {author} {\bibfnamefont {G.~W.}\ \bibnamefont {Kilcup}}, \ and\ \bibinfo {author} {\bibfnamefont {S.~R.}\ \bibnamefont {Sharpe}},\ }\href {\doibase 10.1103/PhysRevD.43.2003} {\bibfield  {journal} {\bibinfo  {journal} {Phys. Rev. D}\ }\textbf {\bibinfo {volume} {43}},\ \bibinfo {pages} {2003} (\bibinfo {year} {1991})}\BibitemShut {NoStop}%
\bibitem [{\citenamefont {Hudspith}(2015)}]{Hudspith:2014oja}%
  \BibitemOpen
  \bibfield  {author} {\bibinfo {author} {\bibfnamefont {R.~J.}\ \bibnamefont {Hudspith}} (\bibinfo {collaboration} {RBC, UKQCD}),\ }\href {\doibase 10.1016/j.cpc.2014.10.017} {\bibfield  {journal} {\bibinfo  {journal} {Comput. Phys. Commun.}\ }\textbf {\bibinfo {volume} {187}},\ \bibinfo {pages} {115} (\bibinfo {year} {2015})},\ \Eprint {http://arxiv.org/abs/1405.5812} {arXiv:1405.5812 [hep-lat]} \BibitemShut {NoStop}%
\bibitem [{\citenamefont {Aubin}\ and\ \citenamefont {Orginos}(2011)}]{Aubin:2010jc}%
  \BibitemOpen
  \bibfield  {author} {\bibinfo {author} {\bibfnamefont {C.}~\bibnamefont {Aubin}}\ and\ \bibinfo {author} {\bibfnamefont {K.}~\bibnamefont {Orginos}},\ }\href {\doibase 10.1063/1.3647217} {\bibfield  {journal} {\bibinfo  {journal} {AIP Conf. Proc.}\ }\textbf {\bibinfo {volume} {1374}},\ \bibinfo {pages} {621} (\bibinfo {year} {2011})},\ \Eprint {http://arxiv.org/abs/1010.0202} {arXiv:1010.0202 [hep-lat]} \BibitemShut {NoStop}%
\bibitem [{\citenamefont {Green}\ \emph {et~al.}(2014)\citenamefont {Green}, \citenamefont {Negele}, \citenamefont {Pochinsky}, \citenamefont {Syritsyn}, \citenamefont {Engelhardt},\ and\ \citenamefont {Krieg}}]{Green:2014xba}%
  \BibitemOpen
  \bibfield  {author} {\bibinfo {author} {\bibfnamefont {J.~R.}\ \bibnamefont {Green}}, \bibinfo {author} {\bibfnamefont {J.~W.}\ \bibnamefont {Negele}}, \bibinfo {author} {\bibfnamefont {A.~V.}\ \bibnamefont {Pochinsky}}, \bibinfo {author} {\bibfnamefont {S.~N.}\ \bibnamefont {Syritsyn}}, \bibinfo {author} {\bibfnamefont {M.}~\bibnamefont {Engelhardt}}, \ and\ \bibinfo {author} {\bibfnamefont {S.}~\bibnamefont {Krieg}},\ }\href {\doibase 10.1103/PhysRevD.90.074507} {\bibfield  {journal} {\bibinfo  {journal} {Phys. Rev. D}\ }\textbf {\bibinfo {volume} {90}},\ \bibinfo {pages} {074507} (\bibinfo {year} {2014})},\ \Eprint {http://arxiv.org/abs/1404.4029} {arXiv:1404.4029 [hep-lat]} \BibitemShut {NoStop}%
\bibitem [{\citenamefont {Fischer}\ \emph {et~al.}(2020)\citenamefont {Fischer}, \citenamefont {Kostrzewa}, \citenamefont {Ostmeyer}, \citenamefont {Ottnad}, \citenamefont {Ueding},\ and\ \citenamefont {Urbach}}]{Fischer:2020bgv}%
  \BibitemOpen
  \bibfield  {author} {\bibinfo {author} {\bibfnamefont {M.}~\bibnamefont {Fischer}}, \bibinfo {author} {\bibfnamefont {B.}~\bibnamefont {Kostrzewa}}, \bibinfo {author} {\bibfnamefont {J.}~\bibnamefont {Ostmeyer}}, \bibinfo {author} {\bibfnamefont {K.}~\bibnamefont {Ottnad}}, \bibinfo {author} {\bibfnamefont {M.}~\bibnamefont {Ueding}}, \ and\ \bibinfo {author} {\bibfnamefont {C.}~\bibnamefont {Urbach}},\ }\href {\doibase 10.1140/epja/s10050-020-00205-w} {\bibfield  {journal} {\bibinfo  {journal} {Eur. Phys. J. A}\ }\textbf {\bibinfo {volume} {56}},\ \bibinfo {pages} {206} (\bibinfo {year} {2020})},\ \Eprint {http://arxiv.org/abs/2004.10472} {arXiv:2004.10472 [hep-lat]} \BibitemShut {NoStop}%
\bibitem [{\citenamefont {Michael}\ and\ \citenamefont {Teasdale}(1983)}]{Michael:1982gb}%
  \BibitemOpen
  \bibfield  {author} {\bibinfo {author} {\bibfnamefont {C.}~\bibnamefont {Michael}}\ and\ \bibinfo {author} {\bibfnamefont {I.}~\bibnamefont {Teasdale}},\ }\href {\doibase 10.1016/0550-3213(83)90674-0} {\bibfield  {journal} {\bibinfo  {journal} {Nucl. Phys. B}\ }\textbf {\bibinfo {volume} {215}},\ \bibinfo {pages} {433} (\bibinfo {year} {1983})}\BibitemShut {NoStop}%
\bibitem [{\citenamefont {L{\"u}scher}\ and\ \citenamefont {Wolff}(1990)}]{Luscher:1990ck}%
  \BibitemOpen
  \bibfield  {author} {\bibinfo {author} {\bibfnamefont {M.}~\bibnamefont {L{\"u}scher}}\ and\ \bibinfo {author} {\bibfnamefont {U.}~\bibnamefont {Wolff}},\ }\href {\doibase 10.1016/0550-3213(90)90540-T} {\bibfield  {journal} {\bibinfo  {journal} {Nucl. Phys. B}\ }\textbf {\bibinfo {volume} {339}},\ \bibinfo {pages} {222} (\bibinfo {year} {1990})}\BibitemShut {NoStop}%
\bibitem [{\citenamefont {Hudspith}\ and\ \citenamefont {Mohler}(2023)}]{Hudspith:2023loy}%
  \BibitemOpen
  \bibfield  {author} {\bibinfo {author} {\bibfnamefont {R.~J.}\ \bibnamefont {Hudspith}}\ and\ \bibinfo {author} {\bibfnamefont {D.}~\bibnamefont {Mohler}},\ }\href {\doibase 10.1103/PhysRevD.107.114510} {\bibfield  {journal} {\bibinfo  {journal} {Phys. Rev. D}\ }\textbf {\bibinfo {volume} {107}},\ \bibinfo {pages} {114510} (\bibinfo {year} {2023})},\ \Eprint {http://arxiv.org/abs/2303.17295} {arXiv:2303.17295 [hep-lat]} \BibitemShut {NoStop}%
\bibitem [{\citenamefont {Semke}\ and\ \citenamefont {Lutz}(2006)}]{Semke:2005sn}%
  \BibitemOpen
  \bibfield  {author} {\bibinfo {author} {\bibfnamefont {A.}~\bibnamefont {Semke}}\ and\ \bibinfo {author} {\bibfnamefont {M.~F.~M.}\ \bibnamefont {Lutz}},\ }\href {\doibase 10.1016/j.nuclphysa.2006.07.043} {\bibfield  {journal} {\bibinfo  {journal} {Nucl.Phys.}\ }\textbf {\bibinfo {volume} {A778}},\ \bibinfo {pages} {153} (\bibinfo {year} {2006})},\ \Eprint {http://arxiv.org/abs/nucl-th/0511061} {arXiv:nucl-th/0511061 [nucl-th]} \BibitemShut {NoStop}%
\bibitem [{\citenamefont {Lutz}\ \emph {et~al.}(2014)\citenamefont {Lutz}, \citenamefont {Bavontaweepanya}, \citenamefont {Kobdaj},\ and\ \citenamefont {Schwarz}}]{Lutz:2014oxa}%
  \BibitemOpen
  \bibfield  {author} {\bibinfo {author} {\bibfnamefont {M.~F.~M.}\ \bibnamefont {Lutz}}, \bibinfo {author} {\bibfnamefont {R.}~\bibnamefont {Bavontaweepanya}}, \bibinfo {author} {\bibfnamefont {C.}~\bibnamefont {Kobdaj}}, \ and\ \bibinfo {author} {\bibfnamefont {K.}~\bibnamefont {Schwarz}},\ }\href {\doibase 10.1103/PhysRevD.90.054505} {\bibfield  {journal} {\bibinfo  {journal} {Phys. Rev.}\ }\textbf {\bibinfo {volume} {D90}},\ \bibinfo {pages} {054505} (\bibinfo {year} {2014})},\ \Eprint {http://arxiv.org/abs/1401.7805} {arXiv:1401.7805 [hep-lat]} \BibitemShut {NoStop}%
\bibitem [{\citenamefont {Lutz}\ \emph {et~al.}(2018)\citenamefont {Lutz}, \citenamefont {Heo},\ and\ \citenamefont {Guo}}]{Lutz:2018cqo}%
  \BibitemOpen
  \bibfield  {author} {\bibinfo {author} {\bibfnamefont {M.~F.~M.}\ \bibnamefont {Lutz}}, \bibinfo {author} {\bibfnamefont {Y.}~\bibnamefont {Heo}}, \ and\ \bibinfo {author} {\bibfnamefont {X.-Y.}\ \bibnamefont {Guo}},\ }\href {\doibase 10.1016/j.nuclphysa.2018.05.007} {\bibfield  {journal} {\bibinfo  {journal} {Nucl. Phys. A}\ }\textbf {\bibinfo {volume} {977}},\ \bibinfo {pages} {146} (\bibinfo {year} {2018})},\ \Eprint {http://arxiv.org/abs/1801.06417} {arXiv:1801.06417 [hep-lat]} \BibitemShut {NoStop}%
\bibitem [{\citenamefont {Guo}\ \emph {et~al.}(2020)\citenamefont {Guo}, \citenamefont {Heo},\ and\ \citenamefont {Lutz}}]{Guo:2019nyp}%
  \BibitemOpen
  \bibfield  {author} {\bibinfo {author} {\bibfnamefont {X.-Y.}\ \bibnamefont {Guo}}, \bibinfo {author} {\bibfnamefont {Y.}~\bibnamefont {Heo}}, \ and\ \bibinfo {author} {\bibfnamefont {M.~F.~M.}\ \bibnamefont {Lutz}},\ }\href {\doibase 10.1140/epjc/s10052-020-7818-9} {\bibfield  {journal} {\bibinfo  {journal} {Eur. Phys. J. C}\ }\textbf {\bibinfo {volume} {80}},\ \bibinfo {pages} {260} (\bibinfo {year} {2020})},\ \Eprint {http://arxiv.org/abs/1907.00714} {arXiv:1907.00714 [hep-lat]} \BibitemShut {NoStop}%
\bibitem [{\citenamefont {Semke}\ and\ \citenamefont {Lutz}(2012)}]{Semke:2011ez}%
  \BibitemOpen
  \bibfield  {author} {\bibinfo {author} {\bibfnamefont {A.}~\bibnamefont {Semke}}\ and\ \bibinfo {author} {\bibfnamefont {M.~F.~M.}\ \bibnamefont {Lutz}},\ }\href {\doibase 10.1103/PhysRevD.85.034001} {\bibfield  {journal} {\bibinfo  {journal} {Phys.Rev.}\ }\textbf {\bibinfo {volume} {D85}},\ \bibinfo {pages} {034001} (\bibinfo {year} {2012})},\ \Eprint {http://arxiv.org/abs/1111.0238} {arXiv:1111.0238 [hep-ph]} \BibitemShut {NoStop}%
\bibitem [{\citenamefont {Tiburzi}\ and\ \citenamefont {Walker-Loud}(2008)}]{Tiburzi:2008bk}%
  \BibitemOpen
  \bibfield  {author} {\bibinfo {author} {\bibfnamefont {B.~C.}\ \bibnamefont {Tiburzi}}\ and\ \bibinfo {author} {\bibfnamefont {A.}~\bibnamefont {Walker-Loud}},\ }\href {\doibase 10.1016/j.physletb.2008.09.054} {\bibfield  {journal} {\bibinfo  {journal} {Phys. Lett. B}\ }\textbf {\bibinfo {volume} {669}},\ \bibinfo {pages} {246} (\bibinfo {year} {2008})},\ \Eprint {http://arxiv.org/abs/0808.0482} {arXiv:0808.0482 [nucl-th]} \BibitemShut {NoStop}%
\bibitem [{\citenamefont {Gell-Mann}\ \emph {et~al.}(1968)\citenamefont {Gell-Mann}, \citenamefont {Oakes},\ and\ \citenamefont {Renner}}]{Gell-Mann:1968hlm}%
  \BibitemOpen
  \bibfield  {author} {\bibinfo {author} {\bibfnamefont {M.}~\bibnamefont {Gell-Mann}}, \bibinfo {author} {\bibfnamefont {R.~J.}\ \bibnamefont {Oakes}}, \ and\ \bibinfo {author} {\bibfnamefont {B.}~\bibnamefont {Renner}},\ }\href {\doibase 10.1103/PhysRev.175.2195} {\bibfield  {journal} {\bibinfo  {journal} {Phys. Rev.}\ }\textbf {\bibinfo {volume} {175}},\ \bibinfo {pages} {2195} (\bibinfo {year} {1968})}\BibitemShut {NoStop}%
\bibitem [{\citenamefont {Gasser}\ and\ \citenamefont {Leutwyler}(1984)}]{Gasser:1983yg}%
  \BibitemOpen
  \bibfield  {author} {\bibinfo {author} {\bibfnamefont {J.}~\bibnamefont {Gasser}}\ and\ \bibinfo {author} {\bibfnamefont {H.}~\bibnamefont {Leutwyler}},\ }\href {\doibase 10.1016/0003-4916(84)90242-2} {\bibfield  {journal} {\bibinfo  {journal} {Annals Phys.}\ }\textbf {\bibinfo {volume} {158}},\ \bibinfo {pages} {142} (\bibinfo {year} {1984})}\BibitemShut {NoStop}%
\bibitem [{\citenamefont {Kuberski}(2024)}]{KUBERSKI2024109173}%
  \BibitemOpen
  \bibfield  {author} {\bibinfo {author} {\bibfnamefont {S.}~\bibnamefont {Kuberski}},\ }\href {\doibase https://doi.org/10.1016/j.cpc.2024.109173} {\bibfield  {journal} {\bibinfo  {journal} {Computer Physics Communications}\ ,\ \bibinfo {pages} {109173}} (\bibinfo {year} {2024})}\BibitemShut {NoStop}%
\bibitem [{\citenamefont {Mohler}\ and\ \citenamefont {Schaefer}(2020)}]{Mohler:2020txx}%
  \BibitemOpen
  \bibfield  {author} {\bibinfo {author} {\bibfnamefont {D.}~\bibnamefont {Mohler}}\ and\ \bibinfo {author} {\bibfnamefont {S.}~\bibnamefont {Schaefer}},\ }\href {\doibase 10.1103/PhysRevD.102.074506} {\bibfield  {journal} {\bibinfo  {journal} {Phys. Rev. D}\ }\textbf {\bibinfo {volume} {102}},\ \bibinfo {pages} {074506} (\bibinfo {year} {2020})},\ \Eprint {http://arxiv.org/abs/2003.13359} {arXiv:2003.13359 [hep-lat]} \BibitemShut {NoStop}%
\bibitem [{\citenamefont {C\`e}\ \emph {et~al.}(2022)\citenamefont {C\`e} \emph {et~al.}}]{Ce:2022kxy}%
  \BibitemOpen
  \bibfield  {author} {\bibinfo {author} {\bibfnamefont {M.}~\bibnamefont {C\`e}} \emph {et~al.},\ }\href {\doibase 10.1103/PhysRevD.106.114502} {\bibfield  {journal} {\bibinfo  {journal} {Phys. Rev. D}\ }\textbf {\bibinfo {volume} {106}},\ \bibinfo {pages} {114502} (\bibinfo {year} {2022})},\ \Eprint {http://arxiv.org/abs/2206.06582} {arXiv:2206.06582 [hep-lat]} \BibitemShut {NoStop}%
\bibitem [{\citenamefont {Aoki}\ \emph {et~al.}(2017)\citenamefont {Aoki} \emph {et~al.}}]{Aoki:2016frl}%
  \BibitemOpen
  \bibfield  {author} {\bibinfo {author} {\bibfnamefont {S.}~\bibnamefont {Aoki}} \emph {et~al.},\ }\href {\doibase 10.1140/epjc/s10052-016-4509-7} {\bibfield  {journal} {\bibinfo  {journal} {Eur. Phys. J. C}\ }\textbf {\bibinfo {volume} {77}},\ \bibinfo {pages} {112} (\bibinfo {year} {2017})},\ \Eprint {http://arxiv.org/abs/1607.00299} {arXiv:1607.00299 [hep-lat]} \BibitemShut {NoStop}%
\bibitem [{\citenamefont {Workman}\ \emph {et~al.}(2022)\citenamefont {Workman} \emph {et~al.}}]{Workman:2022ynf}%
  \BibitemOpen
  \bibfield  {author} {\bibinfo {author} {\bibfnamefont {R.~L.}\ \bibnamefont {Workman}} \emph {et~al.} (\bibinfo {collaboration} {Particle Data Group}),\ }\href {\doibase 10.1093/ptep/ptac097} {\bibfield  {journal} {\bibinfo  {journal} {PTEP}\ }\textbf {\bibinfo {volume} {2022}},\ \bibinfo {pages} {083C01} (\bibinfo {year} {2022})}\BibitemShut {NoStop}%
\bibitem [{\citenamefont {B{\"a}r}\ and\ \citenamefont {Golterman}(2014)}]{Bar:2013ora}%
  \BibitemOpen
  \bibfield  {author} {\bibinfo {author} {\bibfnamefont {O.}~\bibnamefont {B{\"a}r}}\ and\ \bibinfo {author} {\bibfnamefont {M.}~\bibnamefont {Golterman}},\ }\href {\doibase 10.1103/PhysRevD.89.034505} {\bibfield  {journal} {\bibinfo  {journal} {Phys. Rev. D}\ }\textbf {\bibinfo {volume} {89}},\ \bibinfo {pages} {034505} (\bibinfo {year} {2014})},\ \bibinfo {note} {[Erratum: Phys.Rev.D 89, 099905 (2014)]},\ \Eprint {http://arxiv.org/abs/1312.4999} {arXiv:1312.4999 [hep-lat]} \BibitemShut {NoStop}%
\bibitem [{\citenamefont {Chao}\ \emph {et~al.}(2022)\citenamefont {Chao}, \citenamefont {Hudspith}, \citenamefont {G\'erardin}, \citenamefont {Green},\ and\ \citenamefont {Meyer}}]{Chao:2022xzg}%
  \BibitemOpen
  \bibfield  {author} {\bibinfo {author} {\bibfnamefont {E.-H.}\ \bibnamefont {Chao}}, \bibinfo {author} {\bibfnamefont {R.~J.}\ \bibnamefont {Hudspith}}, \bibinfo {author} {\bibfnamefont {A.}~\bibnamefont {G\'erardin}}, \bibinfo {author} {\bibfnamefont {J.~R.}\ \bibnamefont {Green}}, \ and\ \bibinfo {author} {\bibfnamefont {H.~B.}\ \bibnamefont {Meyer}},\ }\href {\doibase 10.1140/epjc/s10052-022-10589-2} {\bibfield  {journal} {\bibinfo  {journal} {Eur. Phys. J. C}\ }\textbf {\bibinfo {volume} {82}},\ \bibinfo {pages} {664} (\bibinfo {year} {2022})},\ \Eprint {http://arxiv.org/abs/2204.08844} {arXiv:2204.08844 [hep-lat]} \BibitemShut {NoStop}%
\bibitem [{\citenamefont {Aoki}\ \emph {et~al.}(2022)\citenamefont {Aoki} \emph {et~al.}}]{FlavourLatticeAveragingGroupFLAG:2021npn}%
  \BibitemOpen
  \bibfield  {author} {\bibinfo {author} {\bibfnamefont {Y.}~\bibnamefont {Aoki}} \emph {et~al.} (\bibinfo {collaboration} {Flavour Lattice Averaging Group (FLAG)}),\ }\href {\doibase 10.1140/epjc/s10052-022-10536-1} {\bibfield  {journal} {\bibinfo  {journal} {Eur. Phys. J. C}\ }\textbf {\bibinfo {volume} {82}},\ \bibinfo {pages} {869} (\bibinfo {year} {2022})},\ \Eprint {http://arxiv.org/abs/2111.09849} {arXiv:2111.09849 [hep-lat]} \BibitemShut {NoStop}%
\bibitem [{\citenamefont {Engel}\ \emph {et~al.}(2013)\citenamefont {Engel}, \citenamefont {Lang}, \citenamefont {Mohler},\ and\ \citenamefont {Sch\"afer}}]{Engel:2013ig}%
  \BibitemOpen
  \bibfield  {author} {\bibinfo {author} {\bibfnamefont {G.~P.}\ \bibnamefont {Engel}}, \bibinfo {author} {\bibfnamefont {C.~B.}\ \bibnamefont {Lang}}, \bibinfo {author} {\bibfnamefont {D.}~\bibnamefont {Mohler}}, \ and\ \bibinfo {author} {\bibfnamefont {A.}~\bibnamefont {Sch\"afer}} (\bibinfo {collaboration} {BGR}),\ }\href {\doibase 10.1103/PhysRevD.87.074504} {\bibfield  {journal} {\bibinfo  {journal} {Phys. Rev. D}\ }\textbf {\bibinfo {volume} {87}},\ \bibinfo {pages} {074504} (\bibinfo {year} {2013})},\ \Eprint {http://arxiv.org/abs/1301.4318} {arXiv:1301.4318 [hep-lat]} \BibitemShut {NoStop}%
\bibitem [{\citenamefont {L{\"u}scher}\ and\ \citenamefont {Schaefer}(2013)}]{Luscher:2012av}%
  \BibitemOpen
  \bibfield  {author} {\bibinfo {author} {\bibfnamefont {M.}~\bibnamefont {L{\"u}scher}}\ and\ \bibinfo {author} {\bibfnamefont {S.}~\bibnamefont {Schaefer}},\ }\href {\doibase 10.1016/j.cpc.2012.10.003} {\bibfield  {journal} {\bibinfo  {journal} {Comput. Phys. Commun.}\ }\textbf {\bibinfo {volume} {184}},\ \bibinfo {pages} {519} (\bibinfo {year} {2013})},\ \Eprint {http://arxiv.org/abs/1206.2809} {arXiv:1206.2809 [hep-lat]} \BibitemShut {NoStop}%
\bibitem [{\citenamefont {Edwards}\ and\ \citenamefont {Joo}(2005)}]{Edwards:2004sx}%
  \BibitemOpen
  \bibfield  {author} {\bibinfo {author} {\bibfnamefont {R.~G.}\ \bibnamefont {Edwards}}\ and\ \bibinfo {author} {\bibfnamefont {B.}~\bibnamefont {Joo}} (\bibinfo {collaboration} {SciDAC, LHPC, UKQCD}),\ }\href {\doibase 10.1016/j.nuclphysbps.2004.11.254} {\bibfield  {journal} {\bibinfo  {journal} {Nucl. Phys. B Proc. Suppl.}\ }\textbf {\bibinfo {volume} {140}},\ \bibinfo {pages} {832} (\bibinfo {year} {2005})},\ \Eprint {http://arxiv.org/abs/hep-lat/0409003} {arXiv:hep-lat/0409003} \BibitemShut {NoStop}%
\bibitem [{\citenamefont {Boyle}\ \emph {et~al.}(2016)\citenamefont {Boyle}, \citenamefont {Cossu}, \citenamefont {Yamaguchi},\ and\ \citenamefont {Portelli}}]{Boyle:2016lbp}%
  \BibitemOpen
  \bibfield  {author} {\bibinfo {author} {\bibfnamefont {P.~A.}\ \bibnamefont {Boyle}}, \bibinfo {author} {\bibfnamefont {G.}~\bibnamefont {Cossu}}, \bibinfo {author} {\bibfnamefont {A.}~\bibnamefont {Yamaguchi}}, \ and\ \bibinfo {author} {\bibfnamefont {A.}~\bibnamefont {Portelli}},\ }\href {\doibase 10.22323/1.251.0023} {\bibfield  {journal} {\bibinfo  {journal} {PoS}\ }\textbf {\bibinfo {volume} {LATTICE2015}},\ \bibinfo {pages} {023} (\bibinfo {year} {2016})}\BibitemShut {NoStop}%
\bibitem [{\citenamefont {Galassi}\ and\ \citenamefont {Al}(2018)}]{Galassi2018}%
  \BibitemOpen
  \bibfield  {author} {\bibinfo {author} {\bibfnamefont {M.}~\bibnamefont {Galassi}}\ and\ \bibinfo {author} {\bibfnamefont {E.}~\bibnamefont {Al}},\ }\href {https://www.gnu.org/software/gsl/} {\emph {\bibinfo {title} {{GNU Scientific Library Reference Manual}}}},\ \bibinfo {edition} {3rd}\ ed.\ (\bibinfo {year} {2018})\BibitemShut {NoStop}%
\end{thebibliography}%
